\documentclass[aps,preprint,showpacs,amsmath,amssymb]{revtex4}
\usepackage{graphicx}
\usepackage{longtable}
\usepackage{dcolumn}  
\usepackage{lscape,ulem}
\usepackage{footnote}
\usepackage{color}
\definecolor{navyblue}{rgb}{0.3,0.3,1}
\definecolor{purple}{rgb}{0.6,0,0.5}
\usepackage[colorlinks=true, pdfstartview=FitV, linkcolor=purple, citecolor=
blue,urlcolor=navyblue]{hyperref}

\begin{document}

\title{Relativistic Mean-Field Hadronic Models under Nuclear Matter 
Constraints} 

\author{M. Dutra} 
\affiliation{Departamento de F\'isica e Matem\'atica - ICT, 
Universidade Federal Fluminense, 28895-532 Rio das Ostras, RJ, Brazil}
\author{O. Louren\c co}
\affiliation{Departamento de Ci\^encias da Natureza, Matem\'atica e 
Educa\c c\~ao, CCA, Universidade Federal de S\~ao Carlos, 13600-970 Araras, SP,
Brazil}
\author{S. S. Avancini}
\affiliation{Depto de F\'{\i}sica - CFM - Universidade Federal de
Santa Catarina, Florian\'opolis - SC - CP. 476 - CEP 88.040 - 900
- Brazil}
\author{B. V. Carlson}
\affiliation{Departamento de F\'isica, Instituto 
Tecnol\'ogico de Aeron\'autica, CTA, 12228-900 S\~ao Jos\'e dos Campos, SP,
Brazil} 
\author{A. Delfino}
\affiliation{Departamento de F\'{\i}sica - Universidade Federal Fluminense,
Av. Litor\^ anea s/n, 24210-150 Boa Viagem, Niter\'oi, RJ, Brazil}
\author{D. P. Menezes} 
\affiliation{Depto de F\'{\i}sica - Universidade Federal de
Santa Catarina, Florian\'opolis - SC - CP. 476 - CEP 88.040 - 900
- Brazil}
\affiliation{Departamento de F\'isica Aplicada, Universidad de Alicante, 
Ap. Correus 99, E-03080, Alicante, Spain}

\author{C.~Provid\^encia}
\affiliation{Centro de F\'isica Computacional, Department of Physics,
University of Coimbra, P-3004-516 Coimbra, Portugal}
\author{S.~Typel}
\affiliation{GSI Helmholtzzentrum f\"ur Schwerionenforschung GmbH, Theorie,
Planckstrasse 1,D-64291 Darmstadt, Germany}
\author{J. R. Stone}
\affiliation{Oxford Physics, University of Oxford, OX1 3PU Oxford, United
Kingdom}
\affiliation{Department of Physics and Astronomy, University of Tennessee, 
Knoxville, Tennessee 37996, USA}

\date{\today}
\begin{abstract}
\noindent{\bf Background:} The microscopic composition and properties of 
infinite hadronic matter at a wide range of densities and temperatures have been 
a subject of intense investigation for decades. The Equation of State (EoS)
relating pressure, energy density and temperature at a given particle
number density is essential for modeling compact astrophysical objects
such as neutron stars, core-collapse supernovae and related phenomena
including the creation of chemical elements in the Universe.  The EoS
depends not only on the particles present in the matter, but, more
importantly, on the forces acting among them. Since a realistic and
quantitative description of infinite hadronic matter and nuclei from
first principles in not available at present, a large variety of
phenomenological models have been developed in the last several
decades, but the scarcity of experimental and observational data does
not allow a unique determination of the adjustable parameters.

\noindent{\bf Purpose:} It is essential for further development of the field 
to determine the most realistic parameter sets and to use them consistently.
Recently a set of constraints on properties of nuclear matter was
formed  and the performance of 240 non-relativistic Skyrme parameterizations 
was assessed \cite{PRC85-035201}, in describing nuclear matter up to about 3 
times nuclear saturation density. In the present work we examine 263 
Relativistic mean-field (RMF) models in a comparable approach. These models have 
been widely used because of several important aspects not always present in 
nonrelativistic models, such as intrinsic Lorentz covariance, automatic 
inclusion of spin, appropriate saturation mechanism for nuclear matter, 
causality and, therefore, no problems related to superluminal speed of sound in 
medium. 

\noindent{\bf Method:} Three different sets of constraints related to 
symmetric nuclear matter, pure neutron matter, symmetry energy, and its 
derivatives were used. The first set (SET1) was the same as used in assessing 
the Skyrme parameterizations. The second and third set (SET2a and SET2b),
were more suitable for analysis of RMF and included, up-to-date
theoretical, experimental and empirical information. 

\noindent{\bf Results:} The sets of updated constraints (SET2a and SET2b) 
differed somewhat in the level of restriction but still yielded only $4$ and 
$3$ approved RMF models, respectively. A similarly small number of approved 
Skyrme parameterizations were found in the previous study with Skyrme models.  
An interesting feature of our analysis has been that the results change 
dramatically if the constraint on the volume part of the isospin 
incompressibility ($K_{\tau,\rm v}$) is eliminated. In this case, we have $35$ 
approved models (SET2a) and $30$ (SET2b). 

\noindent{\bf Conclusions:} Our work provides a new insight into application of 
RMF models to properties of nuclear matter and brings into focus their 
problematic proliferation. The assessment performed in this work should 
be used in future applications of RMF models. Moreover, the most extensive set 
of refined constraints (including nuclear matter and finite-nuclei-related 
properties) should be used in future determinations of new parameter sets in 
order to provide models that can be used with more confidence in a wide range
of applications. Pointing to reasons of the many failures, even of the
frequently used models, should lead to their improvement and to the
identification of possible missing physics not included in present energy 
density functionals.

\end{abstract}

\maketitle

\section{Introduction}

The theoretical description of infinite nuclear matter and finite nuclear properties 
has been relying on models since the primordial developments of nuclear physics. 
Unfortunately, so far there are neither a specific nuclear physics theory nor enough 
adequate solutions for QCD, which is still in the early stages of lattice 
calculations. Many models have been developed since the beginning of the last 
century, from the famous semi-empirical Bethe-Weizs\"acker mass formula proposed 
in 1935 \cite{Wei35}, non-relativistic Skyrme models \cite{skyrme, brink} that 
first appeared around 1950, to relativistic Quantum Hadro Dynamics (QHD) models 
\cite{walecka} developed after 1974, to which we devote our attention in the present 
work.

All relativistic models are written in terms of parameters that are fitted to 
reproduce either bulk nuclear matter or finite nuclei properties. This means that 
most models behave approximately the same as far as equations of state are concerned 
around saturation density and at zero temperature. Nevertheless, these very same 
models have been used to describe physics taking place at sub-saturation
densities, such as liquid-gas phase transitions, and also at very high densities, 
such as neutron star matter. As a consequence, models that describe similar physics 
at saturation density yield very different results when used in the low or high 
density limits. The same holds true if finite systems are investigated. We devote 
the next section to the mean field approximation usually employed when relativistic 
models are considered and to seven different types of parameterizations of QHD {\it 
Walecka-type} models. In these models, the baryons interact among each other by 
exchanging scalar-isoscalar ($\sigma$) and vector-isoscalar ($\omega$) mesons. For 
our analyses of nuclear matter properties, only nucleons are necessary as hadronic 
degrees of freedom. When the models are extended to describe stellar matter, 
hyperons are generally also included. More sophisticated versions include
vector-isovector ($\rho$) and vector-isoscalar ($\delta$) mesons. The seven 
variations we treat next are: 1) the original linear Walecka model, 2) the non-linear 
Walecka model with $\sigma$ self-interacting mesons, 3) the non-linear Walecka model 
with $\sigma$ and $\omega$ self-interacting mesons, 4) the non-linear Walecka model 
with $\sigma$ and $\omega$ self-interacting mesons and possible mesonic cross terms, 
5) models in which the parameters that couple the baryons with the mesons are 
density dependent, 6) point-coupling models, in which the baryons only interact with 
each other through effective point-like interactions, without exchanging mesons, and 
7) models with the inclusion of $\delta$ mesons. Details about the Lagrangian 
density, equations of motion and equations of state for each of the seven model 
types are given in the next section.

In a previous work~\cite{PRC85-035201}, 240 different Skyrme model parameterizations 
were confronted with experimentally and empirically derived constraints and only 16 
of them were shown to satisfy all of the constraints.  The authors argue that the 
production of new parameter sets with a limited range of application should not be
encouraged. In the present work our aim is to obtain the physical properties related 
to the same derived constraints used in~\cite{PRC85-035201} with $263$ relativistic 
models belonging to one of the seven classes mentioned above and check whether they 
satisfy these constraints.

A few words on our choice of constraints are now in order. We start with the isospin 
symmetric nuclear matter incompressibility (or compression modulus) $K_0$, the one 
most used to constrain mean-field models. The incompressibility values have been 
inferred from experiment and from theory. Experimentally, results coming from giant 
resonances, mainly isoscalar giant monopole (GMR)~\cite{pearson} and isovector giant 
dipole (GDR) resonances~\cite{stringari} have been used as an important source of
information. Theoretically, efforts to obtain values for the incompressibility 
started with the use of the Hartree-Fock plus random-phase-approximation 
(RPA)~\cite{blaizot} and continued to other calculations involving even more 
sophisticated treatments. The developments of these calculations can be tracked from 
many papers, but we mention specifically Refs.~\cite{shlomo,PRI-stone}. In the 
present paper, we follow the suggestions given in~\cite{khan1,khan2} for the value of 
$K_0=230 \pm 40$ MeV.

Our second and third most important constraints are the symmetry energy ($J$) and 
its slope ($L_0$) at the saturation density. The density dependence of the symmetry 
energy carries information about the isospin dependence of nuclear forces and gives 
interesting hints on both finite nuclei and neutron star properties. Experimental 
data for the symmetry energy come from various sources, namely heavy-ion 
collisions~\cite{hic}, pygmy dipole resonances~\cite{pdr1,pdr2}, isobaric analog 
states~\cite{ias}, besides GMR and GDR. In~\cite{PRL86-5647}, the authors have shown 
that a direct correlation between the neutron skin thickness (controlled by the 
density dependence of the symmetry energy) and the neutron star radii exists, such 
that models that yield smaller neutron skins in heavy nuclei generate smaller 
neutron star radii. Recent reviews on this subject can be found 
in~\cite{tsang,tamii} and a comprehensive study of the imprint of the symmetry energy 
on the crust and strangeness content of neutron star can be seen in~\cite{review}.
Moreover, in~\cite{PRC66-034305,pieka2004}, a correlation between the values of the
incompressibility and the symmetry energy was proposed based on the fact that the 
isoscalar giant monopole resonance (ISGMR) and isovector giant dipole resonance 
(IVGDR) of $^{208}$Pb were sensitive both to the incompressibility and the symmetry
energy due to its isospin asymmetry. Therefore, the author claims that the ISMGR
data from a nucleus with a well developed breathing mode but a small  neutron-proton 
asymmetry such as $^{90}$Zr should be used to fix  the incompressibility at
saturation instead of a nucleus with a non negligible isospin asymmetry like 
$^{208}$Pb. Once the incompressibility at saturation is fixed, the IVGDR $^{208}$Pb 
may be used to constrain the symmetry energy. Although not conclusive, there seems 
to exist a correlation between the values of $J$, $L_0$ and the curvature ($K_{\rm
sym}^0$) of the symmetry energy at the saturation 
density~\cite{ducoin,lattimerlim,bianca}, and we tackle this point later.

Another constraint that we intend to use is the volume part of the isospin 
incompressibility, known as $K_{\tau,\rm v}$, which depends on several liquid drop 
model quantities. When it is extracted from a simple fitting to GMR data, it 
includes not only volume, but also surface contributions. According to isospin 
diffusion calculations~\cite{chen}, it should be $K_\tau=-500 \pm 50$~MeV. 
According to neutron skin thickness \cite{centelles2009}, 
$K_\tau=-500^{+125}_{-100}$~MeV and according 
to GMR measured in Sn isotopes~\cite{li2007}, $K_\tau=-550 \pm 100$~MeV. In order to 
take into account all these uncertainties, we have chosen $K_\tau=-550 \pm 150$~MeV.

The other constraints are obtained directly from the equation of state or are related to
the constraints mentioned above and we add more comments after they
are defined.

We note that we consider here only systems made of nucleons at zero temperature.
Relativistic models are also used to describe systems at higher densities, including
heavy baryons (hyperons) (see e.g. \cite{lackey2006}), and at finite temperatures,
important in modeling of high density matter in proto-neutron stars and 
core-collapse
supernovae, which we will examine in a separate study.

Our paper is organized as follows. In Sec.~\ref{models} we present the basic 
equations that define the relativistic mean-field models we have chosen to analyze, 
such as the Lagrangian densities and their related equations of state. Our results 
and discussions, including the sets of constraints with which the models are 
confronted, are shown in Sec.~\ref{resdis} and we draw the conclusions in 
Sec.~\ref{conclusions}.

\section{Relativistic Mean-field models at zero temperature}
\label{models}

Relativistic mean-field (RMF) models have been widely used to describe infinite 
nuclear matter (INM), finite nuclei, and stellar matter properties. The main 
representative of such kinds of models, the Walecka model, QHD, or linear Walecka 
model, as it is also known, proposed in 1974~\cite{walecka}, treats protons and 
neutrons as fundamental particles interacting with each other through the exchange 
of scalar and vector mesons. The $\sigma$ and $\omega$ fields represent, 
respectively, these mesons and mimic the attractive and repulsive parts of the 
nuclear interaction.

The two free parameters of the Walecka model, i.e., the couplings between the fields 
and the nucleons, are fitted to reproduce well established properties of infinite 
nuclear matter, namely, the binding energy ($E_0\sim 16$~MeV) and the saturation 
density ($\rho_0\sim 0.15$~fm$^{-3}$). However, it does not give reasonable values 
for the incompressibility ($K_0$), and nucleon effective mass ($M^*$), both related 
to symmetric nuclear matter (SNM). This problem was circumvented by Boguta and 
Bodmer~\cite{boguta}, who added to the Walecka model cubic and quartic self 
interactions in the scalar field $\sigma$, introducing, consequently, two more free 
parameters, which are fitted so as to fix the values of $K_0$ and $M^*$. In the same 
way, more terms can be added to the Boguta-Bodmer model in order to make it 
compatible with other observables, such as those related to finite nuclei. Actually, 
many RMF models and parameterizations have been constructed following this method. 
In our work, we take into account a more general nonlinear finite range RMF model, 
by considering it to be represented by the following Lagrangian density,
\begin{eqnarray}
\mathcal{L}_{\rm NL} = \mathcal{L}_{\rm nm} + \mathcal{L}_\sigma +
\mathcal{L}_\omega
+ \mathcal{L}_\rho + \mathcal{L}_{\delta} + \mathcal{L}_{\sigma\omega\rho},
\label{dl}
\end{eqnarray}
where
\begin{align}
\mathcal{L}_{\rm nm} &= \overline{\psi}(i\gamma^\mu\partial_\mu - M)\psi 
+ g_\sigma\sigma\overline{\psi}\psi 
- g_\omega\overline{\psi}\gamma^\mu\omega_\mu\psi 
- \frac{g_\rho}{2}\overline{\psi}\gamma^\mu\vec{\rho}_\mu\vec{\tau}\psi
+ g_\delta\overline{\psi}\vec{\delta}\vec{\tau}\psi,
\\
\mathcal{L}_\sigma &= \frac{1}{2}(\partial^\mu \sigma \partial_\mu \sigma 
- m^2_\sigma\sigma^2) - \frac{A}{3}\sigma^3 - \frac{B}{4}\sigma^4,
\\
\mathcal{L}_\omega &= -\frac{1}{4}F^{\mu\nu}F_{\mu\nu} 
+ \frac{1}{2}m^2_\omega\omega_\mu\omega^\mu 
+ \frac{C}{4}(g_\omega^2\omega_\mu\omega^\mu)^2,
\\
\mathcal{L}_\rho &= -\frac{1}{4}\vec{B}^{\mu\nu}\vec{B}_{\mu\nu} 
+ \frac{1}{2}m^2_\rho\vec{\rho}_\mu\vec{\rho}^\mu,
\\
\mathcal{L}_\delta &= \frac{1}{2}(\partial^\mu\vec{\delta}\partial_\mu\vec{\delta} 
- m^2_\delta\vec{\delta}^2),
\end{align}
and
\begin{align}
\mathcal{L}_{\sigma\omega\rho} &= 
g_\sigma g_\omega^2\sigma\omega_\mu\omega^\mu
\left(\alpha_1+\frac{1}{2}{\alpha_1}'g_\sigma\sigma\right)
+ g_\sigma g_\rho^2\sigma\vec{\rho}_\mu\vec{\rho}^\mu
\left(\alpha_2+\frac{1}{2}{\alpha_2}'g_\sigma\sigma\right) 
\nonumber \\
&+ \frac{1}{2}{\alpha_3}'g_\omega^2 g_\rho^2\omega_\mu\omega^\mu
\vec{\rho}_\mu\vec{\rho}^\mu.
\label{lomegarho}
\end{align}
In this Lagrangian density, $\mathcal{L}_{\rm nm}$ stands for the kinetic part of 
the nucleons added to the terms representing the interaction between the nucleons 
and mesons $\sigma$, $\delta$, $\omega$, and $\rho$. The term $\mathcal{L}_j$ 
represents the free and self-interacting terms of the meson $j$, for 
$j=\sigma,\delta,\omega,$ and $\rho$. The last term, 
$\mathcal{L}_{\sigma\omega\rho}$, takes into account crossed interactions between 
the meson fields. The antisymmetric field tensors $F_{\mu\nu}$ and 
$\vec{B}_{\mu\nu}$ are given by 
$F_{\mu\nu}=\partial_\nu\omega_\mu-\partial_\mu\omega_\nu$
and $\vec{B}_{\mu\nu}=\partial_\nu\vec{\rho}_\mu-\partial_\mu\vec{\rho}_\nu
- g_\rho (\vec{\rho}_\mu \times \vec{\rho}_\nu)$. The nucleon mass is $M$ and the 
meson masses are $m_j$.

The use of the mean-field approximation, in which the meson fields are treated as
classical fields as
\begin{eqnarray}
\sigma\rightarrow \left<\sigma\right>\equiv\sigma, \quad
\omega_\mu\rightarrow \left<\omega_\mu\right>\equiv\omega_0, \quad
\vec{\rho}_\mu\rightarrow \left<\vec{\rho}_\mu\right>\equiv \bar{\rho}_{0(3)}, \quad
\mbox{and}\quad 
\vec{\delta}\rightarrow\,\,<\vec{\delta}>\equiv\delta_{(3)},
\label{meanfield}
\end{eqnarray}
together with the Euler-Lagrange equations, leads to the following field equations,
\begin{align}
m^2_\sigma\sigma &= g_\sigma\rho_s - A\sigma^2 - B\sigma^3 
+g_\sigma g_\omega^2\omega_0^2(\alpha_1+{\alpha_1}'g_\sigma\sigma)
+g_\sigma g_\rho^2\bar{\rho}_{0(3)}^2(\alpha_2+{\alpha_2}'g_\sigma\sigma)\,\mbox{,}\quad 
\label{sigmaacm}\\
m_\omega^2\omega_0 &= g_\omega\rho - Cg_\omega(g_\omega \omega_0)^3 
- g_\sigma g_\omega^2\sigma\omega_0(2\alpha_1+{\alpha_1}'g_\sigma\sigma)
- {\alpha_3}'g_\omega^2 g_\rho^2\bar{\rho}_{0(3)}^2\omega_0, 
\label{omegaacm}\\
m_\rho^2\bar{\rho}_{0(3)} &= \frac{g_\rho}{2}\rho_3 
-g_\sigma g_\rho^2\sigma\bar{\rho}_{0(3)}(2\alpha_2+{\alpha_2}'g_\sigma\sigma)
-{\alpha_3}'g_\omega^2 g_\rho^2\bar{\rho}_{0(3)}\omega_0^2, 
\label{rhoacm} \\
m_\delta^2\delta_{(3)} &= g_\delta\rho_{s3}, 
\label{deltaacm}\\
[\gamma^\mu (&i\partial_\mu - V_\tau ) - (M+S_\tau)]\psi = 0.
\label{diracacm}
\end{align}
Due to the translational invariance and rotational symmetry of infinite nuclear 
matter, only the zero components of the four-vector fields are nonvanishing. Also 
considering rotational invariance around the third axis in isospin space, we only 
deal with the third components of the isospin space vectors $\vec{\rho}_\mu$ and 
$\vec{\delta}$.

The scalar and vector densities are given by
\begin{align}
\rho_s &=\left<\overline{\psi}\psi\right>={\rho_s}_p+{\rho_s}_n,\quad
\rho_{s3}=\left<\overline{\psi}{\tau}_3\psi\right>={\rho_s}_p-{\rho_s}_n,
\label{rhos}\\
\rho &=\left<\overline{\psi}\gamma^0\psi\right> = \rho_p + \rho_n,\quad
\rho_3=\left<\overline{\psi}\gamma^0{\tau}_3\psi\right> = \rho_p - \rho_n=(2y-1)\rho,
\label{rho}
\end{align}
with 
\begin{eqnarray}
{\rho_s}_{p,n} &=& 
\frac{\gamma M^*_{p,n}}{2\pi^2}\int_0^{{k_F}_{p,n}}
\frac{k^2dk}{\sqrt{k^2+M^{*2}_{p,n}}} 
=\frac{\gamma 
(M^*_{p,n})^3q^2}{2\pi^2}\int_0^1\frac{\xi^2d\xi}{\sqrt{\xi^2+1/q^2}}
\nonumber \\
&=& \frac{\gamma (M^*_{p,n})^{3}}{4\pi^2}\left[q\sqrt{1+q^2}
- \mbox{ln}\left(q+\sqrt{1+q^2}\right)\right],\qquad
\label{rhospn}
\end{eqnarray}
and
\begin{eqnarray}
\rho_{p,n} = \frac{\gamma}{2\pi^2}\int_0^{{k_F}_{p,n}}k^2dk =
\frac{\gamma}{6\pi^2}{k_F^3}_{p,n},
\label{rhopn}
\end{eqnarray}
for $\xi=k/{k_F}_{p,n}$ and $q={k_{F}}_{p,n}/M^{\ast}_{p,n}$ with the indices $p,n$ 
standing for protons and neutrons, respectively. The degeneracy factor is $\gamma=2$ 
for asymmetric matter, and the proton fraction is defined as $y=\rho_p/\rho$. 
The quantity ${k_F}_{p,n}$ is the Fermi momentum in the units in which $\hbar=c=1$.

From the Dirac equation (\ref{diracacm}), we recognize the vector and scalar potentials
written as,
\begin{eqnarray}
V_{\tau\,\mbox{\tiny NL}} &=&g_\omega\omega_0 +
\frac{g_\rho}{2}\bar{\rho}_{0(3)}\tau_3\qquad\mbox{and} 
\label{vectorpotential}\\
S_{\tau\,\mbox{\tiny NL}} &=&-g_\sigma\sigma -g_\delta\delta_{(3)}\tau_3,
\label{scalarpotential}
\end{eqnarray}
with $\tau_3=1$ and $-1$ for protons and neutrons, respectively. We can also define the
effective nucleon mass as $M^*_\tau=M+S_{\tau\,\mbox{\tiny NL}}$, leading to
\begin{eqnarray}
M_p^*=M-g_\sigma\sigma-g_\delta\delta_{(3)} \qquad\mbox{and}\qquad
M_n^*=M-g_\sigma\sigma+g_\delta\delta_{(3)}.
\label{effectivemasses}
\end{eqnarray}
Note the effect of the meson $\delta$, which splits the effective masses $M_p^*$ and 
$M_n^*$. For symmetric nuclear matter $\delta_{(3)}$ vanishes, since 
$\rho_{sp}=\rho_{sn}$, and consequently, $M_p^*=M_n^*=M^*=M-g_\sigma\sigma$.

From the energy-momentum tensor $T^{\mu\nu}$, calculated through the Lagrangian 
density in Eq.~(\ref{dl}), it is possible to obtain the energy density and the 
pressure of the asymmetric system, since $\mathcal{E}=\left<T_{00}\right>$ and 
$P=\left<T_{ii}\right>/3$. These quantities are given as follows,
\begin{eqnarray}
\mathcal{E}_{\rm NL} &=& \frac{1}{2}m^2_\sigma\sigma^2 
+ \frac{A}{3}\sigma^3 + \frac{B}{4}\sigma^4 - \frac{1}{2}m^2_\omega\omega_0^2 
- \frac{C}{4}(g_\omega^2\omega_0^2)^2 - \frac{1}{2}m^2_\rho\bar{\rho}_{0(3)}^2
+g_\omega\omega_0\rho+\frac{g_\rho}{2}\bar{\rho}_{0(3)}\rho_3
\nonumber \\
&+& \frac{1}{2}m^2_\delta\delta^2_{(3)} - g_\sigma g_\omega^2\sigma\omega_0^2
\left(\alpha_1+\frac{1}{2}{\alpha_1}'g_\sigma\sigma\right) 
- g_\sigma g_\rho^2\sigma\bar{\rho}_{0(3)}^2 
\left(\alpha_2+\frac{1}{2}{\alpha_2}' g_\sigma\sigma\right) \nonumber \\
&-& \frac{1}{2}{\alpha_3}'g_\omega^2 g_\rho^2\omega_0^2\bar{\rho}_{0(3)}^2
+ \mathcal{E}_{\mbox{\tiny kin}}^p + \mathcal{E}_{\mbox{\tiny kin}}^n,
\label{denerg}
\end{eqnarray}
where
\begin{eqnarray}
\mathcal{E}_{\mbox{\tiny kin}}^{p,n}&=&\frac{\gamma}{2\pi^2}\int_0^{{k_F}_{p,n}}k^2
(k^2+M^{*2}_{p,n})^{1/2}dk 
\label{decinnlw}= \frac{\gamma {k_F^4}_{p,n}}{2\pi^2}\int_0^1 
\xi^2(\xi^2+z^2)^{1/2}d\xi
\nonumber \\
&=&\frac{\gamma {k_F^4}_{p,n}}{2\pi^2}
\left[\left(1+\frac{z^2}{2}\right)\frac{\sqrt{1+z^2}}{4}
-\frac{z^4}{8}\mbox{ln}\left(\frac{1+\sqrt{1+z^2}}{z}\right)\right]
 \nonumber \\
&=& \frac{3}{4}{E_{F}}_{p,n}\rho_{p,n} + \frac{1}{4}M^{\ast}_{p,n}{\rho_{s}}_{p,n},
\end{eqnarray}
and
\begin{eqnarray}
P_{\rm NL} &=& - \frac{1}{2}m^2_\sigma\sigma^2 - \frac{A}{3}\sigma^3 -
\frac{B}{4}\sigma^4 + \frac{1}{2}m^2_\omega\omega_0^2 
+ \frac{C}{4}(g_\omega^2\omega_0^2)^2 + \frac{1}{2}m^2_\rho\bar{\rho}_{0(3)}^2
+ \frac{1}{2}{\alpha_3}'g_\omega^2 g_\rho^2\omega_0^2\bar{\rho}_{0(3)}^2
\nonumber \\
&-&\frac{1}{2}m^2_\delta\delta^2_{(3)} + g_\sigma g_\omega^2\sigma\omega_0^2
\left(\alpha_1+\frac{1}{2}{\alpha_1}'g_\sigma\sigma\right) 
+ g_\sigma g_\rho^2\sigma\bar{\rho}_{0(3)}^2 
\left(\alpha_2+\frac{1}{2}{\alpha_2}' g_\sigma\sigma\right) \nonumber \\
&+& P_{\mbox{\tiny kin}}^p + P_{\mbox{\tiny kin}}^n,\qquad
\label{pressure}
\end{eqnarray}
with
\begin{eqnarray}
P_{\mbox{\tiny kin}}^{p,n} &=& 
\frac{\gamma}{6\pi^2}\int_0^{{k_F}_{p,n}}\frac{k^4dk}{(k^2+M^{*2}_{p,n})^{1/2}} 
= \frac{\gamma {k_F^4}_{p,n}}{6\pi^2}\int_0^1 \frac{\xi^4d\xi}{(\xi^2+z^2)^{1/2}} 
\nonumber \\
&=&\frac{\gamma {k_F^4}_{p,n}}{6\pi^2}
\left[\left(1-\frac{3z^2}{2}\right)\frac{\sqrt{1+z^2}}{4}
+\frac{3z^4}{8}\mbox{ln}\left(\frac{1+\sqrt{1+z^2}}{z}\right)\right]
\nonumber \\
&=& \frac{1}{4}{E_{F}}_{p,n}\rho_{p,n} - \frac{1}{4}M^{\ast}_{p,n}{\rho_{s}}_{p,n},
\end{eqnarray}
and
\begin{equation}
{E_{F}}_{p,n} = \sqrt{{k_{F}}_{p,n}^{2}+(M^{\ast}_{p,n})^{2}}.
\end{equation}
In the above equations, the parameter $z$ is defined as 
$z=M^{\ast}_{p,n}/{k_{F}}_{p,n}$.

In order to better identify the parameterizations related to the model described in
Eq.~(\ref{dl}), we define here four different types of parameterizations, namely,
\begin{itemize}
 \item {type 1} (linear finite range models): models in which
$A=B=C=\alpha_1=\alpha_2=\alpha_1^\prime=\alpha_2^\prime=\alpha_3^\prime=g_\delta=0$. 
This is the case of the linear Walecka model. Different parameterizations correspond 
to different values of the pair ($\rho_0$,$E_0$).

\item {type 2} ($\sigma^3+\sigma^4$ models): models in which
$C=\alpha_1=\alpha_2=\alpha_1^\prime=\alpha_2^\prime=\alpha_3^\prime=g_\delta=0$. 
This type corresponds to parameterizations related to the Boguta-Bodmer model.

\item {type 3} ($\sigma^3+\sigma^4+\omega_0^4$ models): models in which
$\alpha_1=\alpha_2=\alpha_1^\prime=\alpha_2^\prime=\alpha_3^\prime=g_\delta=0$. 
These parameterizations include a quartic self-interaction in the $\omega$ field.

\item {type 4} ($\sigma^3+\sigma^4+\omega_0^4$ + cross terms models): models in
which $g_\delta=0$ and at least one of the coupling constants, $\alpha_1$, $\alpha_2$,
$\alpha_1^\prime$, $\alpha_2^\prime$, or $\alpha_3^\prime$ is different from zero.

\end{itemize}

Another widely used approach in Quantum Hadrodynamics (QHD) is that in which the 
couplings between nucleons and mesons are sensitive to the nuclear 
medium. Specifically, the RMF model proposed in Ref.~\cite{NPA656-331} allows 
density dependence in the aforementioned couplings by making 
\begin{eqnarray}
g_\sigma\to\Gamma_\sigma(\rho),\quad g_\omega\to\Gamma_\omega(\rho),\quad
g_\rho\to\Gamma_\rho(\rho)\quad\mbox{and}\quad g_\delta\to\Gamma_\delta(\rho). 
\label{ddcouplings}
\end{eqnarray}
Its Lagrangian density is given by
\begin{eqnarray}
\mathcal{L}_{\rm DD} &=& \overline{\psi}(i\gamma^\mu\partial_\mu - M)\psi 
+ \Gamma_\sigma(\rho)\sigma\overline{\psi}\psi 
- \Gamma_\omega(\rho)\overline{\psi}\gamma^\mu\omega_\mu\psi 
-\frac{\Gamma_\rho(\rho)}{2}\overline{\psi}\gamma^\mu\vec{\rho}_\mu\vec{\tau}
\psi + \Gamma_\delta(\rho)\overline{\psi}\vec{\delta}\vec{\tau}\psi \nonumber \\
&+& \frac{1}{2}(\partial^\mu \sigma \partial_\mu \sigma - m^2_\sigma\sigma^2)
- \frac{1}{4}F^{\mu\nu}F_{\mu\nu} + \frac{1}{2}m^2_\omega\omega_\mu\omega^\mu 
-\frac{1}{4}\vec{B}^{\mu\nu}\vec{B}_{\mu\nu}+\frac{1}{2}m^2_\rho
\vec{\rho}_\mu\vec{\rho}^\mu \nonumber \\
&+& \frac{1}{2}(\partial^\mu\vec{\delta}\partial_\mu\vec{\delta} 
- m^2_\delta\vec{\delta}^2),
\label{dldd}
\end{eqnarray}
where
\begin{eqnarray}
\Gamma_i(\rho) &=& \Gamma_i(\rho_0)f_i(x),\quad\mbox{with}\quad
f_i(x) = a_i\frac{1+b_i(x+d_i)^2}{1+c_i(x+d_i)^2},
\label{gamadefault}
\end{eqnarray}
for $i=\sigma,\omega$, and
\begin{eqnarray}
\Gamma_\rho(\rho)=\Gamma_\rho(\rho_0)e^{-a(x-1)},\quad\mbox{with}\quad x=\rho/\rho_0.
\end{eqnarray}

Some density dependent parameterizations have couplings different from those of the 
above equations. In particular, the GDFM model~\cite{PRC77-025802} presents the 
following form for its couplings,
\begin{eqnarray}
\Gamma_i(\rho)=a_i+(b_i+d_ix^3)e^{-c_ix},
\end{eqnarray}
for $i=\sigma,\omega,\rho,\delta$. A correction to the 
coupling parameter for
the meson $\omega$  is also taken into account,
\begin{eqnarray}
\Gamma_{\mbox{\tiny cor}}(\rho)=\Gamma_\omega(\rho) - a_{\mbox{\tiny
cor}}e^{-\left(\frac{\rho-\rho_0}{b_{\mbox{\tiny cor}}}\right)^2}.
\end{eqnarray}
The DDH$\delta$ parameterization has the same coupling parameters as 
in Eq.~(\ref{gamadefault}) for the mesons $\sigma$ and $\omega$, but functions 
$f_i(x)$ given
by \cite{ddhdelta}
\begin{eqnarray}
f_i(x)=a_ie^{-b_i(x-1)}-c_i(x-d_i),
\end{eqnarray}
for $i=\rho,\delta$.

By applying the mean-field approximation and the Euler-Lagrange equations, we find the
same field equations as in Eqs.~(\ref{sigmaacm})-(\ref{diracacm}), taking into 
account Eq.~(\ref{ddcouplings}) and 
$A=B=C=\alpha_1=\alpha_2=\alpha_1^\prime=\alpha_2^\prime=\alpha_3^\prime=0$. The 
scalar and vector densities are defined as in the previous nonlinear RMF model. 
The proton and neutron effective masses, $M^*_p$ and $M^*_n$, and the scalar 
potential are also defined as in Eqs.~(\ref{effectivemasses}) and 
(\ref{scalarpotential}), respectively, observing the generalizations in 
Eq.~(\ref{ddcouplings}). The same does not occur for the vector potential that now 
reads,
\begin{eqnarray}
V_{\tau\,\mbox{\tiny DD}}&=&\Gamma_\omega(\rho)\omega_0 +
\frac{\Gamma_\rho(\rho)}{2}\bar{\rho}_{0(3)}\tau_3 + \Sigma_R(\rho),
\end{eqnarray}
with
\begin{eqnarray}
\Sigma_R(\rho)=\frac{\partial\Gamma_\omega}{\partial\rho}\omega_0\rho
+\frac{1}{2}\frac{\partial\Gamma_\rho}{\partial\rho}\bar{\rho}_{0(3)}\rho_3
-\frac{\partial\Gamma_\sigma}{\partial\rho}\sigma\rho_s
-\frac{\partial\Gamma_\delta}{\partial\rho}\delta_{(3)}\rho_{s3}
\end{eqnarray}
being the rearrangement term.

The energy density and pressure are given, respectively, by
\begin{eqnarray}
\mathcal{E}_{\rm DD} &=& \frac{1}{2}m^2_\sigma\sigma^2 
- \frac{1}{2}m^2_\omega\omega_0^2 
- \frac{1}{2}m^2_\rho\bar{\rho}_{0(3)}^2
+ \frac{1}{2}m^2_\delta\delta^2_{(3)}
+\Gamma_\omega(\rho)\omega_0\rho+\frac{\Gamma_\rho(\rho)}{2}\bar{\rho}_{0(3)}\rho_3
\nonumber \\
&+& \mathcal{E}_{\mbox{\tiny kin}}^p + \mathcal{E}_{\mbox{\tiny kin}}^n,
\label{denergdd} \qquad\mbox{and}
\\
P_{\rm DD} &=& \rho\Sigma_R(\rho)- \frac{1}{2}m^2_\sigma\sigma^2 +
\frac{1}{2}m^2_\omega\omega_0^2 
+ \frac{1}{2}m^2_\rho\bar{\rho}_{0(3)}^2
-\frac{1}{2}m^2_\delta\delta^2_{(3)} 
+ P_{\mbox{\tiny kin}}^p + P_{\mbox{\tiny kin}}^n.
\label{pressuredd}
\end{eqnarray}

We also define here the fifth type of parameterization analyzed in our work, namely,
\begin{itemize}
\item {type 5} (density dependent models): parameterizations obtained from
Eq.~(\ref{dldd}) in which $\Gamma_\delta=0$.
\end{itemize}

Another class of RMF models is the nonlinear point-coupling (NLPC) 
model~\cite{PRC46-1757}. In this theory, nucleons interact with each other only 
through effective point-like interactions, without exchanging mesons. It can be 
easily proved that the linear version of the PC model results in the same equations 
of state as the linear Walecka model~\cite{lpc}. The same does not hold for NLPC 
models with cubic and quartic interactions, and finite range RMF models of type~$2$, 
besides both versions can describe equally well infinite nuclear matter~\cite{prc}. 
Here, we  treat a general type of NLPC model, described by the following 
Lagrangian density,
\begin{eqnarray}
\mathcal{L}_{\rm NLPC} &=& \overline{\psi}(i\gamma^\mu\partial_\mu - M)\psi 
-\frac{\alpha_s}{2}(\overline{\psi}\psi)^2
-\frac{\beta_s}{3}(\overline{\psi}\psi)^3
-\frac{\gamma_s}{4}(\overline{\psi}\psi)^4
-\frac{\alpha_{\mbox{\tiny V}}}{2}(\overline{\psi}\gamma^\mu\psi)^2
-\frac{\gamma_{\mbox{\tiny V}}}{4}(\overline{\psi}\gamma^\mu\psi)^4 \nonumber \\
&-&\frac{\alpha_{\mbox{\tiny TV}}}{2}(\overline{\psi}\gamma^\mu\vec{\tau}\psi)^2
-\frac{\gamma_{\mbox{\tiny TV}}}{4}(\overline{\psi}\gamma^\mu\vec{\tau}\psi)^4
-\frac{\alpha_{\mbox{\tiny TS}}}{2}(\overline{\psi}\vec{\tau}\psi)^2
+\left[\eta_1+\eta_2(\overline{\psi}\psi)\right](\overline{\psi}\psi)(\overline{\psi}
\gamma^\mu\psi)^2 \nonumber \\
&-&\eta_3(\overline{\psi}\psi)(\overline{\psi}\gamma^\mu\vec{\tau}\psi)^2.
\label{dlnlpc}
\end{eqnarray}

The mean-field approximation gives rise to the Dirac equation given 
in Eq.~(\ref{diracacm}), with the vector and scalar potentials modified, 
respectively, as follows,
\begin{eqnarray}
V_{\tau\,\mbox{\tiny NLPC}}
&=&\alpha_{\mbox{\tiny V}}\rho+\alpha_{\mbox{\tiny
TV}}\rho_3\tau_3+\gamma_{\mbox{\tiny V}}\rho^3+\gamma_{\mbox{\tiny TV}}
\rho^3_3\tau_3+2(\eta_1+\eta_2\rho_s)\rho_s\rho+2\eta_3\rho_s\rho_3\tau_3, \\
S_{\tau\,\mbox{\tiny NLPC}}
&=&\alpha_s\rho_s+\beta_s\rho_s^2+\gamma_s\rho_s^3+\eta_1\rho^2+2\eta_2\rho_s\rho^2+\eta_3
\rho_3^2++\alpha_{\mbox{\tiny TS}}\rho_{s3}\tau_3,
\end{eqnarray}
with $\rho_s$, $\rho_{s3}$, $\rho$ and $\rho_3$ defined as in
Eqs.~(\ref{rhos})-(\ref{rhopn}). The proton and neutron effective masses read,
\begin{eqnarray}
M_p^*=M+\alpha_s\rho_s+\beta_s\rho_s^2+\gamma_s\rho_s^3+\eta_1\rho^2+2\eta_2\rho_s\rho^2
+\eta_ 3\rho_3^2+\alpha_{\mbox{\tiny TS}}\rho_{s3}, \\
M_n^*=M+\alpha_s\rho_s+\beta_s\rho_s^2+\gamma_s\rho_s^3+\eta_1\rho^2+2\eta_2\rho_s\rho^2
+\eta_ 3\rho_3^2-\alpha_{\mbox{\tiny TS}}\rho_{s3}.
\end{eqnarray}
Notice that the interaction whose magnitude is given by $\alpha_{\mbox{\tiny TS}}$ plays
the same role as the meson $\delta$ in the finite range RMF models, namely, the splitting
of proton and neutron effective masses.

Finally, we obtain the energy density and the pressure for the NLPC model as,
\begin{eqnarray}
\mathcal{E}_{\rm NLPC} &=& \frac{1}{2}\alpha_{\mbox{\tiny V}}\rho^2 +
\frac{1}{2}\alpha_{\mbox{\tiny TV}}\rho_3^2 
+\frac{1}{4}\gamma_{\mbox{\tiny V}}\rho^4 + \frac{1}{4}\gamma_{\mbox{\tiny TV}}\rho_3^4 -
\eta_2\rho_s^2\rho^2 - \frac{1}{2}\alpha_s\rho_s^2 - \frac{2}{3}\beta_s\rho_s^3 -
\frac{3}{4}\gamma_s\rho_s^4 
\nonumber \\
&-& \frac{1}{2}\alpha_{\mbox{\tiny TS}}\rho_{s3}^2 + \mathcal{E}_{\mbox{\tiny kin}}^p +
\mathcal{E}_{\mbox{\tiny kin}}^n,\label{denergnlpc} \qquad\mbox{and}\\
P_{\rm NLPC} &=& \frac{1}{2}\alpha_{\mbox{\tiny V}}\rho^2 +
\frac{1}{2}\alpha_{\mbox{\tiny TV}}\rho_3^2 
+\frac{3}{4}\gamma_{\mbox{\tiny V}}\rho^4 + \frac{3}{4}\gamma_{\mbox{\tiny TV}}\rho_3^4
+2\eta_1\rho_s\rho^2 
+ 3\eta_2\rho_s^2\rho^2 + 2\eta_3\rho_s\rho_3^2 + \frac{1}{2}\alpha_s\rho_s^2 
\nonumber \\
&+& \frac{2}{3}\beta_s\rho_s^3 + \frac{3}{4}\gamma_s\rho_s^4
+\frac{1}{2}\alpha_{\mbox{\tiny TS}}\rho_{s3}^2 + P_{\mbox{\tiny kin}}^p + P_{\mbox{\tiny
kin}}^n,
\label{pressurenlpc}
\end{eqnarray}
respectively. We now define the sixth type of the analyzed parameterizations
as,
\begin{itemize}
\item {type 6} (point-coupling models): parameterizations of the model described by
Eq.~(\ref{dlnlpc}) in which $\alpha_{\mbox{\tiny TS}}=0$.
\end{itemize}

For the sake of completeness, we still define a last type of parameterizations as,
\begin{itemize}
\item {type 7} (delta meson models): parameterizations of finite range models
presenting the meson~$\delta$, i.e., models in which $g_\delta\neq 0$ in the respective
Lagrangian density and equations of state.
\end{itemize}

We still calculate the symmetry energy, $\mathcal{S}(\rho)$, for the RMF
model shown here from the general definition,
\begin{eqnarray}
\mathcal{S}(\rho) &=& \frac{1}{8}\frac{\partial^2(\mathcal{E}/\rho)}{\partial
y^2}\bigg|_{\rho,y=1/2}.
\end{eqnarray}
The expressions are
\begin{eqnarray}
\mathcal{S}_{\rm NL}(\rho) &=& \frac{k_F^2}{6E_F^*}
+ \frac{g_\rho^2}{8{m_\rho^*}^2}\rho -
\left(\frac{g_\delta}{m_\delta}\right)^2\frac{M^{*2}\rho}{2E_F^{*2}[
1+(g_\delta/m_\delta)^2A( k_F,M^*)]},
\label{esym} \\
\mathcal{S}_{\rm DD}(\rho) &=& \frac{k_F^2}{6E_F^*}
+ \frac{\Gamma_\rho^2}{8{m_\rho}^2}\rho -
\left(\frac{\Gamma_\delta}{m_\delta}\right)^2\frac{M^{*2}\rho}{2E_F^{*2}[
1+(\Gamma_\delta/m_\delta)^2A( k_F,M^*)]},
\label{esymdd} \\
\mathcal{S}_{\rm NLPC}(\rho) &=& \frac{k_F^2}{6E_F^*} + \frac{1}{2}\alpha_{\mbox{\tiny
TV}}\rho + \eta_3\rho_s\rho + \frac{1}{2}\alpha_{\mbox{\tiny TS}}
\frac{M^{*2}\rho}{2E_F^{*2}[1-\alpha_{\mbox{\tiny TS}}A(k_F,M^*)]},
\end{eqnarray}
with
\begin{eqnarray}
E_F^*&=&(k_F^2+{M^*}^2)^{1/2}, \\
A(k_F,M^*)&=&\frac{2}{\pi^2}\int_0^{k_F}\frac{k^4dk}{(k^2+M^{*2})^{3/2}},\quad\mbox{and}\\
{m_\rho^*}^2&=&m_\rho^2 + g_\sigma g_\rho^2\sigma(2\alpha_2+\alpha_2' g_\sigma\sigma)
+\alpha_3'g_\omega^2g_\rho^2\omega_0^2.
\end{eqnarray}

All quantities chosen as constraints that are presented in the next section are
directly calculated from $P$, $\mathcal{E}$ and $\mathcal{S}$, as can
be seen from the following definitions,
\begin{eqnarray}
K_0 &=& 9\left(\frac{\partial
P}{\partial\rho}\right)_{\rho=\rho_0,y=1/2},\quad\mbox{\small (incompressibility)}\\
Q_0&=&27\rho_0^3\frac{\partial^3(\mathcal{E}/\rho)}{\partial\rho^3}
\bigg|_{\rho=\rho_0,y=1/2}, \quad\mbox{\small (skewness coefficient)}\\
J&=&\mathcal{S}(\rho_0),\quad\mbox{\small (symmetry energy at $\rho=\rho_0$)} \\
L_0&=&3\rho_0\left(\frac{\partial\mathcal{S}}{\partial\rho}\right)_{\rho=\rho_0},
\quad\mbox{\small (slope of $\mathcal{S}$)}\\
K_{\rm sym}^0&=&9\rho_0^2\left(\frac{\partial^2\mathcal{S}}{\partial\rho^2}\right)
_{\rho=\rho_0}, \quad\mbox{\small (curvature of $\mathcal{S}$)}\\
Q_{\rm sym}^0 &=&
27\rho_0^3\left(\frac{\partial^3\mathcal{S}}{\partial\rho^3}\right)
_{\rho=\rho_0}, \quad\mbox{\small (skewness of $\mathcal{S}$)}\\
K_{\tau,\rm v}^0 &=& \left(K^0_{\rm sym} - 6L_0 -
\frac{Q_0}{K_0}L_0\right).\quad\mbox{\small (volume part of the isospin 
incompressibility)},
\label{kat0v}
\end{eqnarray}
where $\rho_0$ is the saturation density, see Ref.~\cite{vidana}.

In Appendix~\ref{type4models} we explicitly calculate some of the quantities defined 
above as a function of density for RMF models of types $4$, $5$ and $6$ at zero 
temperature. In the finite temperature regime, the integrals in $\rho_{p,n}$, 
${\rho_s}_{p,n}$, $\mathcal{E}_{\mbox{\tiny kin}}^{p,n}$ and $P_{\mbox{\tiny 
kin}}^{p,n}$ should be replaced by those extending from zero to infinity and taking 
into account the Fermi-Dirac distributions for particles and 
anti-particles~\cite{silva}.
 
The saturation properties of all $263$ RMF parameterizations analyzed in this work 
are displayed in Table~\ref{tab:sat} of Appendix~\ref{apsatprop}, with the 
corresponding references.

\section{Results and discussions}
\label{resdis}

\subsection{Previous constraints: SET1}

We next present two sets of constraints used to analyse $263$ relativistic models. 
First, we take the same $11$ constraints previously used to analyse the Skyrme-type
parameterizations. Each individual constraint is explained in detail also 
in Ref.~\cite{PRC85-035201}. Here, we summarize them by stressing 
that they are closely related to properties {\it (i) of symmetric nuclear matter 
(SNM)}: the incompressibility~(SM1), skewness coefficient~(SM2), density dependence 
of pressure in the ranges of \mbox{$2<\frac{\rho}{\rho_0}<4.6$}~(SM3) and 
\mbox{$1.2<\frac{\rho}{\rho_0}<2.2$}~(SM4), {\it (ii) of pure neutron matter (PNM)}: 
the density dependence of energy per particle~(PNM1) and pressure~(PNM2), {\it (iii) 
involving both SNM and PNM}: the symmetry energy~(MIX1), its slope~(MIX2) and the 
volume part of the isospin incompressibility~(MIX3), all of them evaluated at the 
saturation density, the ratio of the symmetry energy at $\rho_0/2$ to its value at 
$\rho_0$~(MIX4), and the ratio of $3P_{\mbox{\tiny PNM}}(\rho_0)$ to 
$L_0\rho_0$~(MIX5). This set of constraints, named here as {SET1}, is shown 
in Table~\ref{tab:set1} with its respective range of validity.
\begin{table}[!htb]
\scriptsize
\begin{ruledtabular}
\caption{List of the macroscopic constraints of SET1 and the range of their 
experimental/empirical (exp/emp) values, density region in which they are valid
and the corresponding range obtained using the approved RMF models (CRMF).}
\centering
\begin{tabular}{lcccccc}
Constraint & Quantity      & Density Region       & Range of constraint  
& Range of constraint      & Ref.\\ 
&           &       & (exp/emp)
& from CRMF         & \\ \hline
SM1   & K$_0$ & $\rho_0$ (fm$^{-3}$) & 200 $-$ 260 MeV & 271.0
MeV & \cite{PRL82-691}\\
SM2   & K$^\prime=-$Q$_0$ & $\rho_0$ (fm$^{-3}$) & 200 $-$ 1200
MeV& 733.6 MeV & \cite{NPA615-135}\\
SM3   & P($\rho$)   & $2<\frac{\rho}{\rho_0}<4.6$ & Band Region & 
see Fig.~\ref{fig:set1}    & \cite{SCI298-1592}\\
SM4   & P($\rho$)  & $1.2<\frac{\rho}{\rho_0}<2.2$    & Band Region  & see
Fig.~\ref{fig:set1}  & \cite{PPNP62-427}\\
PNM1  & $\mathcal{E}_{\mbox{\tiny PNM}}/\rho$ & $0.017<\frac{\rho}{\rho_{\rm
o}}<0.108$   & Band Region &  see Fig.~\ref{fig:set1} &
\cite{PRC85-035201}\\
PNM2  & P($\rho$) & $2<\frac{\rho}{\rho_0}<4.6$ & Band Region    & see
Fig.~\ref{fig:set1}   & \cite{SCI298-1592}\\
MIX1  & $J$ & $\rho_0$ (fm$^{-3}$) & 30 $-$ 35 MeV  & 33.8 $-$ 34.0 MeV
& \cite{PPNP58-587}\\ 
MIX2  & $L_0$ & $\rho_0$ (fm$^{-3}$) & 40 $-$ 76 MeV & 70.9 $-$ 73.9 MeV 
 & \cite{PRC82-024321}\\
MIX3 & $K_{\rm \tau,v}^0$ & $\rho_0$ (fm$^{-3}$) & -760 $-$ -372 MeV &
-388.5 $-$ -388.4 MeV   & \cite{PRI-stone} \\
MIX4  & $\frac{\mathcal{S}(\rho_0/2)}{J}$  & $\rho_0$ (fm$^{-3}$)
& 0.57 $-$ 0.86 & 0.58  & \cite{NPA727-233}\\
MIX5  & $\frac{3P_{\mbox{\tiny PNM}}}{L_0\rho_0}$ & $\rho_0$ (fm$^{-3}$) &
0.90 $-$ 1.10 & 1.05 $-$ 1.06    & \cite{PRC76-064310}\\
\end{tabular}
\label{tab:set1}
\end{ruledtabular}
\end{table}

\begin{figure}[!htb]
\begin{center}
\includegraphics[scale=0.24]{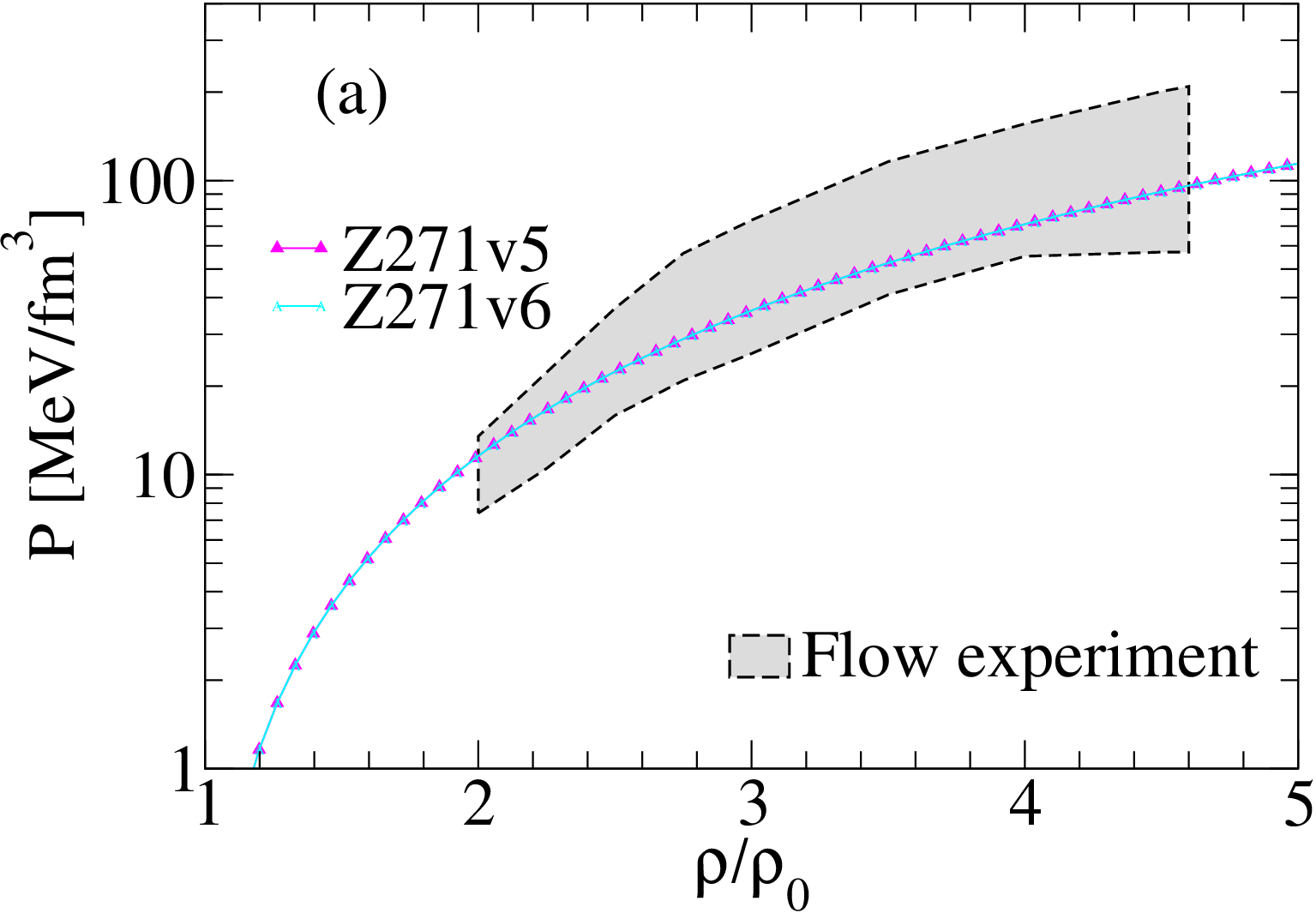}\\
\includegraphics[scale=0.24]{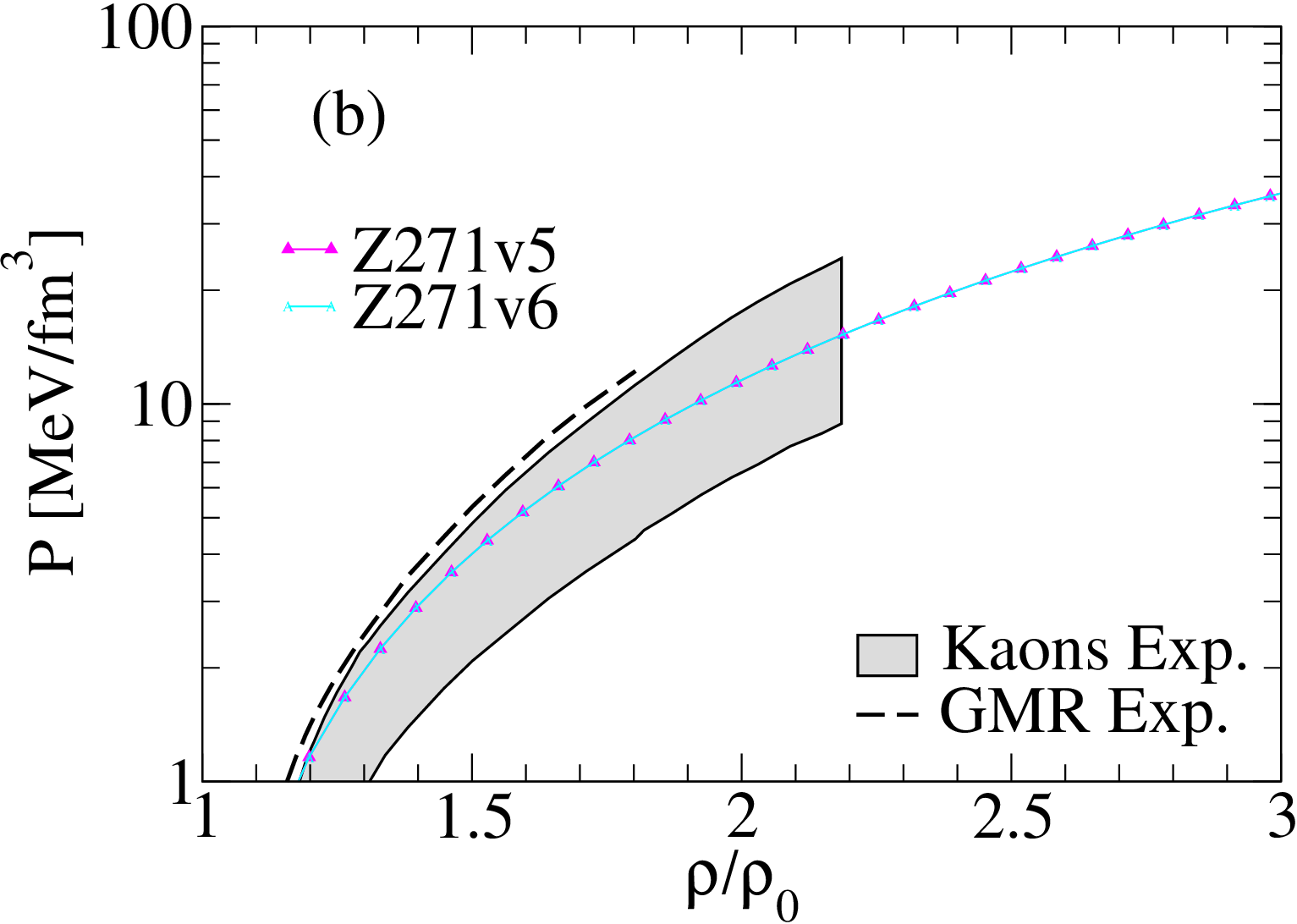}\\
\hspace{0.6cm}\includegraphics[scale=0.24]{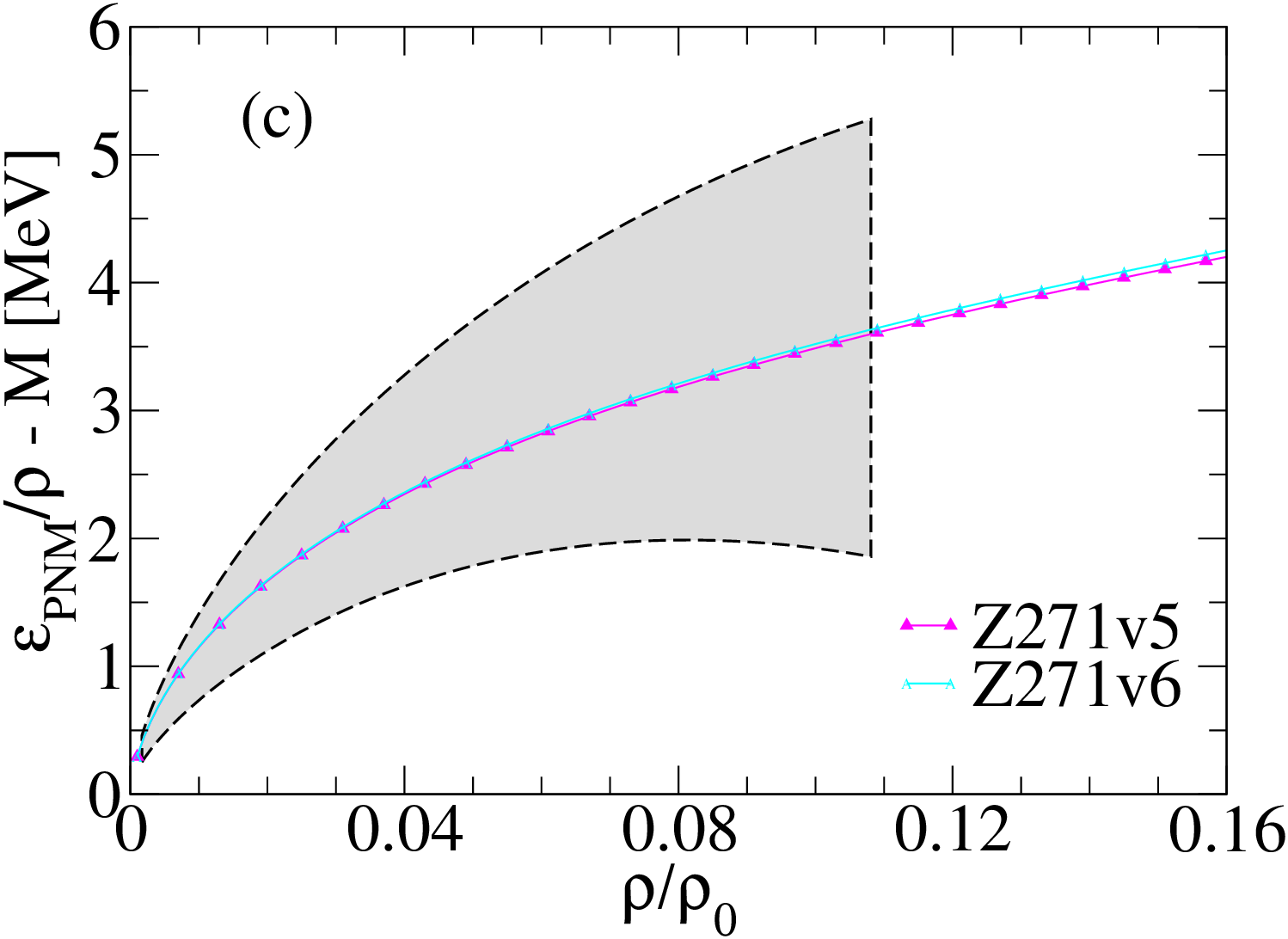}\\
\includegraphics[scale=0.24]{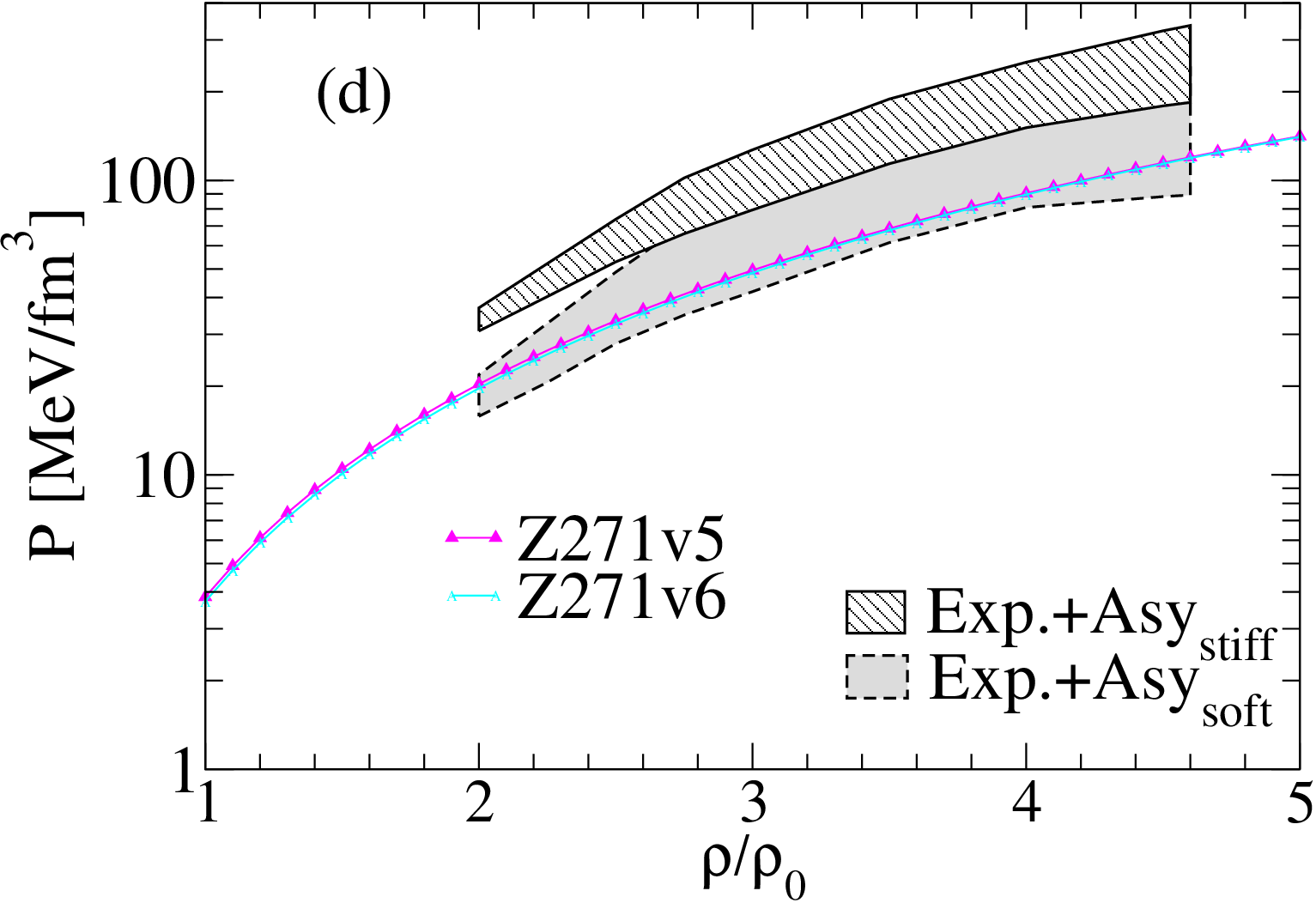}
\caption{(Color online) Density dependence of the approved models Z271v5 and Z271v6 in 
the a) SM3, b) SM4, c) PNM1 and d) PNM2 constraints related to SET1. The shaded bands 
were extracted from Ref.~\cite{SCI298-1592}, where flow experimental data is 
compared with results obtained for a) symmetric matter and d) pure neutron 
matter, b) Ref.~\cite{PPNP62-427}, where pressure in symmetric matter is 
compared with data extracted from kaon production, c) Ref.~\cite{PRC85-035201}, 
as explained in the text.}
\label{fig:set1}
\end{center}
\end{figure}

Regarding specifically pure neutron matter, the related EoS is of a 
particular interest, because PNM is a realistic first approximation to the 
baryonic matter that composes neutron stars. Most properties of neutron stars 
can not be studied in terrestrial laboratories and theoretical models, based on 
effective forces, must be used. However, at low densities, experiments with 
cold Fermi atoms yield information on strongly interaction fluids, very similar 
to the low-density neutron matter at neutron star crusts~\cite{gezerlis2008}. 
Different density regimes can be tuned by the magnitude of the neutron Fermi 
momentum $k_{\rm F}$ relative to the effective range $r_{\rm o}$ of the NN 
interaction in the system~\cite{PRL95-160401}. The ground state energy per 
particle can be expressed as $E_{\rm PNM}/E_{\rm PNM}^{o} = \xi$,
where $E_{\rm PNM} = {\mathcal E_{\rm PNM}}/{\rho}$ and $E_{\rm 
PNM}^{o}$ is the kinetic part of $E_{\rm PNM}$. In the dilute degenerate Fermi 
gas regime ($k_F r_{\rm o} \ll 1$), $\xi$ is a constant~\cite{carlson2003a}. 
This restricts the density below about $10^{\rm -3}\rho_{\rm o}$, the density 
at which neutrons become unbound in neutron stars. At higher densities, below 
$\sim 0.1\rho_{\rm o}$, where  $k_Fr_{\rm o} \approx 1$, $\xi$ has to be 
replaced by a system dependent function $\xi(k_{\rm F},r_{\rm o})$. Likewise 
in Ref.~\cite{PRC85-035201}, we adopt here the expression 
$E_{\rm PNM}/E_{\rm PNM}^{o}$ by Epelbaum {\it et al.}~\cite{EPJA40-199}, 
based on next-to-leading order in lattice chiral effective field theory 
(NLO$_{\rm 3}$), and including corrections due to finite scattering length, 
nonzero effective
range, and higher order corrections,
\begin{eqnarray}
\frac{E_{PNM}}{E_{PNM}^{o}} = \xi-\frac{\xi_1}{k_Fa_{\rm o}} + c_1k_Fr_{\rm o} +
c_2k_F^2m_\pi^{-2} + c_3k_F^3m_\pi^{-3}+\cdots, 
\label{eq:106} 
\end{eqnarray}
where $m_{\rm \pi}$ is the pion mass. The dimensionless universal constant $\xi$
has been determined from trapped cold atom experiments with $^{\rm 6}$Li and
$^{\rm 40}$K, which yield a variety of values: 
$0.32_{-13}^{+10}$~\cite{bartenstein2004}, $0.51(4)$~\cite{kinast2005}, 
$0.46_{-05}^{+12}$~\cite{stewart2006}, and $0.39(2)$~\cite{luo2008}. Values of 
$\xi_1$ in the literature are in the range $0.8 - 1.0$~(\cite{EPJA40-199} and 
references therein). Epelbaum {\it et al.}, using a simple Hamiltonian and only 
few particles in their system, took  $\xi=0.31$ and $\xi_{1}=0.81$ and fitted 
two sets of constants, namely, ($c_1=0.27, c_2=-0.44, c_3=0.0$) and ($c_1=0.17, 
c_2=0.0, c_3=-0.26$), and obtained a very similar quality fits to their 
NLO$_{\rm 3}$. We used the constraint on energy per particle of PNM in
the range of densities $0.01 - 0.1\rho_{\rm o}$ showed in Figs.~\ref{fig:set1}c 
and~\ref{fig:set2}c with $\xi_{1}=0.81$ and the two sets ($c_1,c_2,c_3$). The 
band is obtained by taking $0.2 < \xi < 0.6$, which allows for the spread in 
experimental values. This is the constraint named as PNM1.

As in Ref.~\cite{PRC85-035201}, a model is considered approved in a numerical 
constraint if its deviation, given as 
\begin{eqnarray}
\mbox{Dev} =\frac{Q_{\mbox{\tiny mod}}-Q_{\mbox{\tiny const}}}{\Delta},
\label{dev}
\end{eqnarray}
obeys the relation $|\mbox{Dev}|\leqslant 1$ with $Q_{\mbox{\tiny mod}}$ being the 
value of the quantity calculated in the model, $Q_{\mbox{\tiny const}}$ the central 
value of the related constraint, and $\Delta$ the error related to $Q_{\mbox{\tiny 
const}}$. Specifically for the MIX1, MIX3 and MIX4 constraints, we define their 
central values as $Q_{\mbox{\tiny const}} = (x_2 + x_1)/2$ and the error as 
$\Delta=x_2 - Q_{\mbox{\tiny const}} = Q_{\mbox{\tiny const}} - x_1$, since they are 
given in the form of $x_1\leqslant X\leqslant x_2$. On the other hand, a graphic 
constraint is satisfied if the model is inside the corresponding band in $95\%$ or 
more of the density region.

We also present in the fifth column of Table~\ref{tab:set1}, the range of the 
quantities used in each constraint obtained from the approved models which are
designated by Consistent Relativistic Mean-Field (CRMF). This range is defined from 
the smaller and larger values of the respective quantity chosen among the selected 
models. For instance, the range of the slope of the symmetry energy, $70.9\leqslant 
L_0\leqslant 73.9$, is constructed by noting that $L_0^{\mbox{\tiny 
Z271v6}}=70.9$~MeV and $L_0^{\mbox{\tiny Z271v5}}=73.9$~MeV. For the graphic 
constraints, we analyze the density dependence of the Z271v5 and Z271v6 
parameterizations in Fig.~\ref{fig:set1}.

By applying the constraints of SET1 to the RMF parameterizations, we obtain the
result that {\it none of the models} satisfies all constraints simultaneously. 
However, there are $12$ parameterizations that do not satisfy only one individual 
constraint, i.e., they are consistent with the $10$ remaining ones. We specify such 
parametrizations in Table~\ref{tab:only1}, including the constraint they do not satisfy, 
and the respective deviation obtained from Eq.~(\ref{dev}).
\begin{table}[!htb]
\scriptsize
\begin{ruledtabular}
\caption{List of parametrizations that fail in only one constraint of SET1.}
\centering
\begin{tabular}{lcc}
Model & Model value (MeV) & Deviation \\
\hline
\multicolumn{2}{c}\mbox{constraint not satisfied: SM1 ($200\leqslant K_0 \leqslant 260$ 
MeV)}\\
\hline
Z271v5		&   $271.00$  & $1.37$ \\
Z271v6		&   $271.00$  & $1.37$ \\
\hline
\multicolumn{2}{c}\mbox{constraint not satisfied: MIX3 ($-760\leqslant K_{\rm \tau,v}^0 
\leqslant -372$ MeV)}\\
\hline
BSR15		&  -$252.54$  & $1.62$ \\
BSR16		&  -$258.75$  & $1.58$ \\
FSUGold		&  -$276.07$  & $1.49$ \\ 
FSUGZ06		&  -$259.47$  & $1.58$ \\
FSUGold4	&  -$205.59$  & $1.86$ \\
FSU-III		&  -$341.03$  & $1.16$ \\
FSU-IV		&  -$210.68$  & $1.83$ \\
TW99		&  -$332.32$  & $1.20$ \\
DD-F		&  -$285.54$  & $1.45$ \\
DD-ME$\delta$	&  -$258.28$  & $1.59$ \\
\end{tabular}
\label{tab:only1}
\end{ruledtabular}
\end{table}

In these $12$ parameterizations, two specific ones, models of type~$4$ (cross terms 
models), namely, {Z271v5} and {Z271v6}, fall outside the limits of the SM1 
constraint (the one not satisfied only for these two models) by less then $5\%$, i. e., 
their incompressibilities, $K_0^{\mbox{\tiny 271v5}}=K_0^{\mbox{\tiny 271v6}}=271$~MeV, 
exceed the highest value of the SM1 constraint, $K_{0,\mbox{\tiny SM1}}^{\mbox{\tiny 
max}}=260$~MeV, by $11$~MeV and the ratio of this excess to $K_{0,\mbox{\tiny 
SM1}}=230$~MeV is less than $5\%$. For such cases, we apply the same criterium of 
Ref.~\cite{PRC85-035201} and define these models as included in the CRMF 
parameterizations, \mbox{i. e.}, the ones satisfying the constraints of SET1. 

For the sake of completeness, we provide in Table~\ref{tab:napset1} the number 
of the RMF approved models for each constraint of SET1.
\begin{table}[!htb]
\begin{ruledtabular}
\caption{Number of approved models (among $263$) in each constraint of SET1.}
\centering
\begin{tabular}{lccccccccccc}
Constraints  & SM1   & SM2   & SM3   & SM4   & PNM1  & PNM2 & MIX1  & MIX2   &
MIX3  & MIX4  & MIX5 \\
\hline
Number of models      & $146$ & $174$ & $104$ & $153$ & $193$ & $101$ & $162$ 
& $59$ & $124$ & $65$ & $258$ \\
\end{tabular}
\label{tab:napset1}
\end{ruledtabular}
\end{table}

It is worth noting that the MIX5 constraint is the ``weakest'' one, while the 
constraint defined by the slope of the symmetry energy, MIX2, is the ``strongest'' 
among all of them, since only $58$ of $263$ parameterizations present $L_0$ in the 
range of $L_0=58\pm18$~MeV.

\subsection{Updated constraints: SET2a and SET2b}

It is well known that some of the validity ranges are different if obtained 
for Skyrme-type or relativistic models. Several studies involving both RMF models 
and non-relativistic models have shown that although both sets of models verify the same 
correlations, the parameter distribution of each set is not completely overlapping, see, 
for instance, Refs.~\cite{pdr2,vidana,centelles2009}. This is a consequence of the 
different structure of these models, namely, the existence of a scalar density completely 
absent in the non-relativistic models which gives contributions corresponding to 
many-body effects~\cite{brito}. In fact, the constraints imposed within each of the sets 
are not measured directly, but they result from the analysis of raw data, which involves model 
assumptions. As an example, we refer that the incompressibility derived from GMR using 
Skyrme interactions is around $230$~MeV, but many of the RMF models predict higher values 
(see e.g. Ref.~\cite{PRI-stone}). In a similar way, a value $J=32.5\pm 0.5$~MeV is 
found by fitting a large set of experimental data in the Finite-Range-Droplet-Model 
(FRDM)~\cite{frdm}. The extrapolation of the various fits for the non-relativistic 
models (Skyrme and Gogny) yield typical values of $J$ in the region of $27$ to $38$~MeV. 
Also for such a quantity, the various RMF parameterizations yield higher values, see 
Fig.~18 of Ref.~\cite{PPNP58-587} for a clear comparison. Based on this phenomenology, 
we have opted for other sets of constraints that are somewhat more adequate to analyse 
the RMF models. In these new sets, named hereafter as {SET2a} and {SET2b}, we 
also intend to take into account new theoretical, experimental and empirical information 
concerning the quantities related to the constraints used in this work.

First of all we discuss the SM1 constraint, related to the incompressibility of 
infinity nuclear matter. In the recent study of Refs.~\cite{khan1,khan2}, the authors 
investigated the density dependence of the incompressibility in various relativistic 
and nonrelativistic models, finding a crossing point around a density of 
$\rho_c=0.7\rho_0$. They pointed out the existence of this crossing also in other 
bulk properties such as the symmetry energy  \cite{ducoin} and the energy density of 
pure neutron matter \cite{brown00}. In fact, phenomenological models are generally 
fitted to finite nuclei which provide fitting constraints at a  density slightly 
below saturation, making $\rho_c$ more suitable to characterize nuclear finite 
systems. In this perspective, the authors have shown that the quantity 
$M_c=3\rho_c\frac{\partial K(\rho)}{\partial\rho}\big|_{\rho=\rho_c}$ is more 
strongly correlated with $E_{\mbox{\tiny GMR}}$ than $K_0=K(\rho_0)$, and claimed to 
use this relation to first constrain $M_c$ and after to infer the value of $K_0$ 
($E_{\mbox{\tiny GMR}}$ is the centroid energy of the isoscalar giant monopole 
resonance). This is an alternative to the often used method of constraining $K_0$ 
directly from its correlation with $E_{\mbox{\tiny GMR}}$. Following this new 
approach, and by using experimental data of $E_{\mbox{\tiny GMR}}$ for $^{208}\rm 
Pb$, $^{112-124}\rm Sn$ isotopes, $^{90}\rm Zr$, and $^{144}\rm Sm$, the linear 
correlation of $E_{\mbox{\tiny GMR}}$ and $M_c$ shown in Refs.~\cite{khan1,khan2} was
used to constrain $M_c$ to the range $M_c=1100\pm70$~MeV. Thereby, a new range of
$K_0$ was proposed by noting that $K_0$ and $M_c$ are also linearly 
correlated~\cite{khan1,khan2}. In our new set of updated constraints, we will use this
range of values for $K_0$ as our new {SM1} constraint, in this case given by 
$K_0=230\pm 40$~MeV. Notice that this new constraint is slightly less restrictive 
than the old one given by \mbox{$K_0=230\pm30$~MeV}. 

Although interesting, we note that the new constraint is based on theory and a specific
selection of Skyrme forces (with the exception of FSUGold, DDME2 and D1S which seem to
cluster just at the low $M_c$ values). The selection of experimental $E_{\mbox{\tiny
GMR}}$ is also limited, there are many more values $E_{\mbox{\tiny GMR}}$ in the
literature which would have to be used to verify the conclusion of Fig.~3 (Fig.~2) in
Ref.~\cite{khan1} (Ref.~\cite{khan2}). For example, there are three different sets of the
experimental $E_{\mbox{\tiny GMR}}$ in $^{208}$Pb which differ outside errors: $13.91\pm
0.11$~MeV, $13.90\pm 0.30$~MeV, $14.24$~MeV, $14.17\pm 0.28$~MeV, $14.18\pm 0.11$~MeV (see
discussion in Ref.~\cite{youngblood1999}), $13.96 \pm 0.20$~MeV \cite{youngblood2004},
$13.4\pm 0.2$~MeV \cite{uchida2004} and $13.5\pm 0.2$~MeV \cite{uchida2003}. If the lowest
value, $13.2$~MeV, allowed by the error was used, the constraint on $M_c$ would be
different. Also, the error on $E_{\mbox{\tiny GMR}}$ $^{120}$Sn is $200$~keV, not
$100$~keV ($15.4 \pm 0.2$~MeV) \cite{li2007,li2010}.

Another constraint directly established from a correlation with the incompressibility 
is the SM2 one. In Ref.~\cite{NPA615-135}, the authors used the leptodermous 
expansion for the incompressibility of a finite nucleus of mass number $A$ and 
radius $R$, 
\begin{eqnarray}
K(A,y)& = & K_0 + K_{\rm surf}A^{-1/3} + K_{\rm curv}A^{-2/3}
+K_{\tau,\rm v}^0(1-2y)^2 + K_{\rm coul} \frac{Z^{\rm 2}}{A^{\rm 4/3}} + \cdots,
\label{kay}
\end{eqnarray}
and its relation with $E_{\mbox{\tiny GMR}}$ through
\begin{eqnarray}
K(A,y)=(M/\hbar^{\rm 2})\left<R^2\right>E^2_{\mbox{\tiny GMR}}, 
\end{eqnarray}
to find the range of $K^\prime=700\pm 500$~MeV (the volume part of the isospin 
incompressibility, $K_{\tau,\rm v}^0$, is written in terms of $K^\prime=-Q_0$ in
Eq.~(\ref{kat0v})). This constraint was obtained from the range predicted for the
incompressibility, $K_0=215\pm 15$~MeV, extracted from the comparison of experimental 
values of $E_{\mbox{\tiny GMR}}$ and those calculated from theoretical models. 
However, the analysis of Ref.~\cite{NPA615-135} was entirely based on
parameterizations of the nonrelativistic Skyrme model. For this reason, and because 
of the difference between $K_0$ found in Ref.~\cite{NPA615-135} and the one defined 
in our new SM1 constraint, we have decided here to eliminate the SM2 constraint from 
SET2 and, consequently, from the updated analysis of the RMF parameterizations.

The SM3 constraint is related to the limits of the density dependence of the pressure 
of infinite nuclear matter. In the previous work~\cite{PRC85-035201}, we taken these 
limits as defined by Danielewicz and co-authors in Ref.~\cite{SCI298-1592}. In that 
work, the authors established such limits from analysis of transverse and elliptical 
flows of the ejected particles in the $^{197}\rm Au$ + $^{197}\rm Au$
collisions. 

In a recent study~\cite{steiner2012}, Steiner {\it et al.}, extracted the radius $r$ of a
$1.4$ solar mass neutron star in the range $10.4\leqslant r\leqslant 12.9$~km, generating
a new constraint that equations of state must satisfy for the mass-radius relation of
neutron stars. Their analysis was based on observational data of (i) bursting neutron
stars showing photospheric radius expansion, and of (ii) transiently accreting neutron
stars in quiescence. As a consequence, the authors also established a new range of
validity for the density dependence of the pressure of infinity nuclear matter, consistent
with the previous SM3 constraint proposed in Ref.~\cite{SCI298-1592} in the lower pressure
region. In the high pressure region, however, the new constraint is broader than the
former. In order to take into account this new phenomenology, we use here the SM3
constraint in two different levels, namely, the {SM3a}, in which we consider the band
in the density dependence of pressure increased by $20\%$ in its upper limit, and the
{SM3b}, in which we use the band exactly as in the SM3 constraint of SET1. It is 
worth noting that the SM3a constraint is a less restrictive version of the SM3 one 
used in SET1. 

We note that very recently the radius of a $1.4$ solar mass neutron star
\cite{steiner2012} has been updated by Lattimer and Steiner \cite{lattimer2014} to be
$11.15$ - $12.66$~km (with $95\%$ confidence) if a nucleon-only equation of state is
considered and $10.45$ - $12.45$ km for an equation of state with {\it exotic} components.
Their analysis was performed on exactly the same five objects as the analysis reported by
Guillot {\it et al.}  \cite{guillot2013}, who found the radius of a wide range of neutron
star masses to be $9.1^{+1.3}_{-1.5}$~km (with $90\%$ confidence). Further observation and
analysis would be desirable to refine constraints related to mass-radius of cold neutron
stars.

In Ref.~\cite{SCI298-1592}, Danielewicz {\it et al.}, also proposed a constraint on 
the density dependence of the pressure in pure neutron matter. They made an 
extrapolation of data concerning transversal and elliptical flow of the previous 
case by including asymmetry terms with strong and weak density dependence in the 
pressure. This constraint, named PNM2, was used in Ref.~\cite{PRC85-035201} and also 
in SET1. However, in SET2 we decide not to consider it, by understanding that no new 
experimental information on heavy-ion collisions is present in such an 
extrapolation. Therefore, the PNM2 constraint is absent from SET2.

\begin{figure}[!htb]
\begin{center}
\includegraphics[scale=0.5]{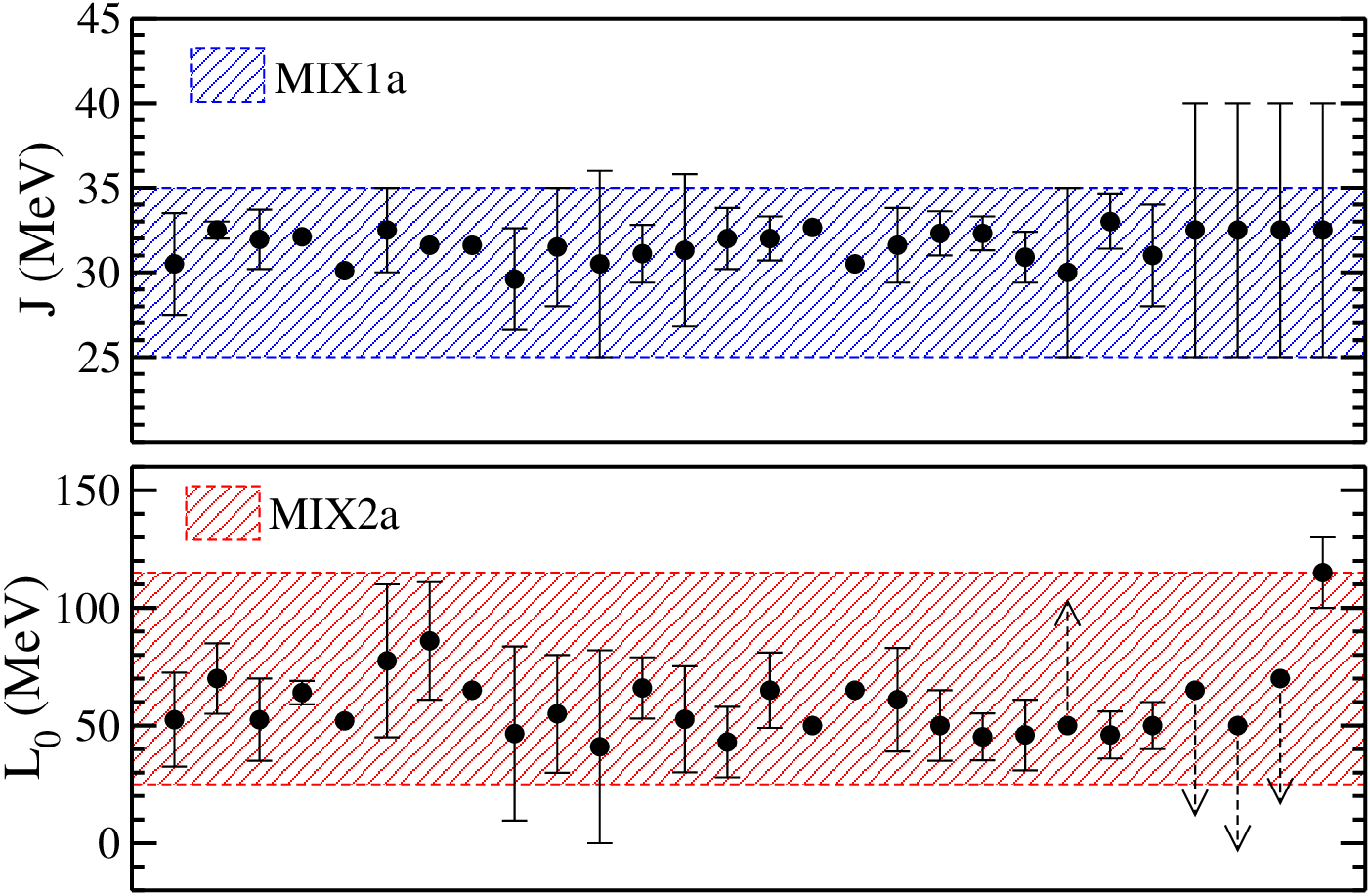}
\caption{(Color online) Comparison between the limits used in MIX1a and MIX2a 
constraints, and those from $28$ different experimental/observational data collected in 
Ref~\cite{jlvalues}.}
\label{figjl}
\end{center}
\end{figure}

Concerning the constraints on the symmetry energy at the saturation density ($J$), we
decide to consider in SET2 the fact that effective hadronic models present values for $J$
in a broader range than that used in SET1, namely, $30\leqslant J\leqslant 35$~MeV. We
consider here a small modification in the lower limit of this range, giving rise to 
our {MIX1a} constraint in which $25\leqslant J\leqslant 35$~MeV. By the same 
token, we modify the range of the slope of the symmetry energy at $\rho_0$ ($L_0$), 
to the new one given by $25\leqslant L_0\leqslant 115$~MeV. This constraint is named 
{MIX2a}. On the other hand, we call the reader's attention to the
fact that the new limits established for MIX1a and MIX2a are totally compatible 
with experimental values available in the literature. In order to make this clear, we 
present in Fig.~\ref{figjl}, a set of twenty-eight $J$ and $L$ values extracted from 
Ref.~\cite{jlvalues}, in which the authors collected from the literature data obtained 
from analyses of different terrestrial nuclear experiments and astrophysical 
observations. They include analyses of isospin diffusion, neutron skins, pygmy dipole
resonances, $\alpha$ and $\beta$ decays, transverse flow, the mass-radius relation of 
neutron stars, and torsional crust oscillations of neutron stars. As one can see 
in Fig.~\ref{figjl}, the new constraints encompass all experimental/observational data.

The reader is reminded that this comparison, based on searching for an overlap of 
the outcome of many experimental/observational methods, inherently implies that all
of them have the same weight, which may not be consensus in the literature. However,
we wish to make clear that this comparison has been used in this work only as a
guideline, in order to take into account the more recent data regarding the possible
ranges of validity for $J$ and $L$.

We also compare the broader range of $J$ with our previous one of 
Ref.~\cite{PRC85-035201}, named in SET2 as {MIX1b}, and, for the sake of completeness, we 
also compare the new range of $L_0$ with a combination of the ranges recently given in 
Refs.~\cite{tsang,baoanli}. We name this more restrictive range for $L_0$, 
namely, $30\leqslant L_0\leqslant 80$~MeV, as {MIX2b}.

To finish the list of updated constraints, we make several remarks concerning the 
limits of the quantity $K_{\tau,\rm v}^0$. First of all, we point out here that 
such a quantity is extremely relevant in our study, since it represents the volume 
part of the isospin incompressibility. It is important to analyze the values that RMF 
models predict for this observable, in the sense that such an investigation can offer
a clue for the improvement of the isospin part of effective RMF interactions. According 
to the literature, there are at least three methods for finding a constraint on the value 
of $K_\tau$, defined as $K_\tau=K_{\tau,\rm v}^0+K_{\tau,\rm s}$, where the last term is 
related to the surface part of the isospin incompressibility. In Ref.~\cite{chen}, the 
authors used an isospin and momentum dependent transport model, to obtain the degree of 
isospin diffusion in the $^{124}\rm Sn+{}^{112}\rm Sn$, $^{124}\rm Sn+{}^{124}\rm Sn$ and 
$^{112}\rm Sn+{}^{112}\rm Sn$ collisions, at energies of $50$~MeV/nucleon and impact 
parameter of $6$~fm. The correlation between this degree of diffusion and the 
isospin incompressibility, led to the constraint of $-500 \pm 50$~MeV for $K_\tau$. 
On the other hand, Centelles {\it et al.} found in Ref.~\cite{centelles2009} that 
$c_{\mbox{\tiny sym}}(\rho\simeq 0.1\mbox{fm}^{-3})=a_{\mbox{\tiny sym}}(A)$, where 
$c_{\mbox{\tiny sym}}(\rho)\simeq J-L_0\epsilon+\frac{1}{2}K^0_{\mbox{\tiny 
sym}}\epsilon^2$ (with $\epsilon=\frac{\rho_0-\rho}{3\rho_0}$), and the symmetry 
energy coefficient of finite nuclei is $a_{\mbox{\tiny 
sym}}(A)=J/[1+(9J/4\mathcal{Q})A^{1/3}]$ ($\mathcal{Q}$ is the surface
stiffness). From this relation, and using that $a_{\mbox{\tiny sym}}(A)$ is a linear 
function of the neutron skin thickness, the authors estimated a range of 
$K_\tau=-500^{+125}_{-100}$~MeV, by analyzing the neutron skin thickness of
$26$ antiprotonic atoms. Finally, in Ref.~\cite{li2007}, the authors used an 
expression for $K(A,y)$ similar to Eq. (\ref{kay}) to obtain the value of 
$K_\tau=-550\pm 100$~MeV from data of $E_{\mbox{\tiny GMR}}$. 

Regarding the theoretical calculations for $K_\tau$, it is also important to mention 
here that mean-field models predict only its volume term, i.e., $K_{\tau,\rm  v}^0$. 
However, as it is rather tricky to separate the volume from the surface term (although it has 
been done several times, see e.g. Ref.~\cite{PRI-stone} and references therein), and 
as the volume term seems to be dominant, we have opted to constrain the calculated 
volume term with $K_\tau$ values obtained from experiment, i.e. we have assumed $K_{\tau 
\rm, v}^0\approx K_\tau$ for the RMF models. Finally, in order to take into account all 
aforementioned procedures for the estimation of $K_\tau$, we define here our new 
constraint {MIX3} as $K_\tau=-550\pm 150$~MeV.

Finally, we stress here that the MIX5 constraint was also removed from our new 
analysis because it is a very ``weak'' constraint, in the sense that practically all 
models are approved in its range, see Table~\ref{tab:napset1}. Thus, we do not 
consider it as a good model selector.

The list of the updated constraints together with those not modified from SET1 is 
given in Table~\ref{tab:set2}.
\begin{table}[hb!]
\scriptsize
\begin{ruledtabular}
\caption{The same as in Table~\ref{tab:set1} now taking into account the following 
updated constraints: SM1, SM3a, SM3b, MIX1a, MIX1b, MIX2a, MIX2b and MIX3. The SM4, 
PNM1 and MIX4 constraints are the same as in SET1. The SM2, PNM2 and MIX5 
constraints were removed. These new constraints are used to generate two new sets, 
namely, SET2a and SET2b (see the text for their definitions).}
\centering
\begin{tabular}{lcccccc}
Constraint & Quantity      & Density Region       & Range of constraint  
& Range of constraint      & Ref.\\ 
&           &       & (exp/emp)
& from CRMF         & \\ \hline
SM1   & K$_0$ & $\rho_0$ (fm$^{-3}$) & 190 $-$ 270 MeV & 225.2
$-$ 232.4 MeV & \cite{khan1,khan2}\\
SM3a  & \multicolumn{4}{c}{the same as SM3b plus $20\%$ on upper
  limit \hspace{2.1cm} see Fig.~\ref{fig:set2}} & \cite{steiner2012}\\
SM3b   & P($\rho$)   & $2<\frac{\rho}{\rho_0}<5$ & Band Region & 
see Fig.~\ref{fig:set2}    & \cite{SCI298-1592}\\
SM4   & P($\rho$)  & $1.2<\frac{\rho}{\rho_0}<2.2$    & Band Region  & see
Fig.~\ref{fig:set2}  & \cite{PPNP62-427}\\
PNM1  & $\mathcal{E}_{\mbox{\tiny PNM}}/\rho$ & $0.017<\frac{\rho}{\rho_{\rm
o}}<0.108$   & Band Region &  see Fig.~\ref{fig:set2} &
\cite{PRC85-035201}\\
MIX1a  & $J$ & $\rho_0$ (fm$^{-3}$) & 25 $-$ 35 MeV  & 33.2 $-$ 34.2 MeV  
& \\
MIX1b  & $J$ & $\rho_0$ (fm$^{-3}$) & 30 $-$ 35 MeV  & 33.2 $-$ 34.0 MeV  
& \cite{PPNP58-587}\\
MIX2a  & $L_0$ & $\rho_0$ (fm$^{-3}$) & 25 $-$ 115 MeV & 77.9 $-$ 84.8 MeV
& \\
MIX2b  & $L_0$ & $\rho_0$ (fm$^{-3}$) & 30 $-$ 80 MeV & 77.9 $-$ 78.8 MeV   
& \cite{tsang,baoanli}\\
MIX3 & $K_{\rm \tau,v}^0$ & $\rho_0$ (fm$^{-3}$) & -700 $-$ -400 MeV &
-421.6(a)/-414.3(b) $-$ -382.5 MeV   &  \\
MIX4  & $\frac{\mathcal{S}(\rho_0/2)}{J}$  & $\rho_0$ (fm$^{-3}$)
& 0.57 $-$ 0.86 & 0.57(a)/0.59(b) $-$ 0.59 & \cite{NPA727-233}\\
\end{tabular}
\label{tab:set2}
\end{ruledtabular}
\end{table}

From these updated constraints we define two distinct sets. One of them composed by 
the constraints in which their ranges are broader, {SET2a}, and other one in 
which they are more stringent, namely, {SET2b}. Specifically, SET2a (SET2b) is 
defined by the SM1, SM3a (SM3b), SM4, PNM1, MIX1a (MIX1b), MIX2a (MIX2b), MIX3 and 
MIX4 constraints.

The application of the constraints of SET2a to the $263$ RMF parameterizations 
collected in our work leads to only {\it 2 models} satisfying all constraints 
simultaneously. They are the type 4 models {BSR12} and {BKA24}. In addition, 
$24$ parameterizations satisfy all constraints of SET2a except one, i. e., they are 
approved in $7$ out of the $8$ constraints. In this group, the {BSR11} and {BKA22} models, 
also of type 4, fell outside the range of the unsatisfied constraint 
(in this case, MIX3) by less than $5\%$. Therefore, as we have done in 
Ref.~\cite{PRC85-035201}, and in the previous analysis of SET1, we consider these 
{\it 4 models} as belonging to the CRMF models of SET2a.
\begin{figure}[!htb]
\begin{center}
\includegraphics[scale=0.24]{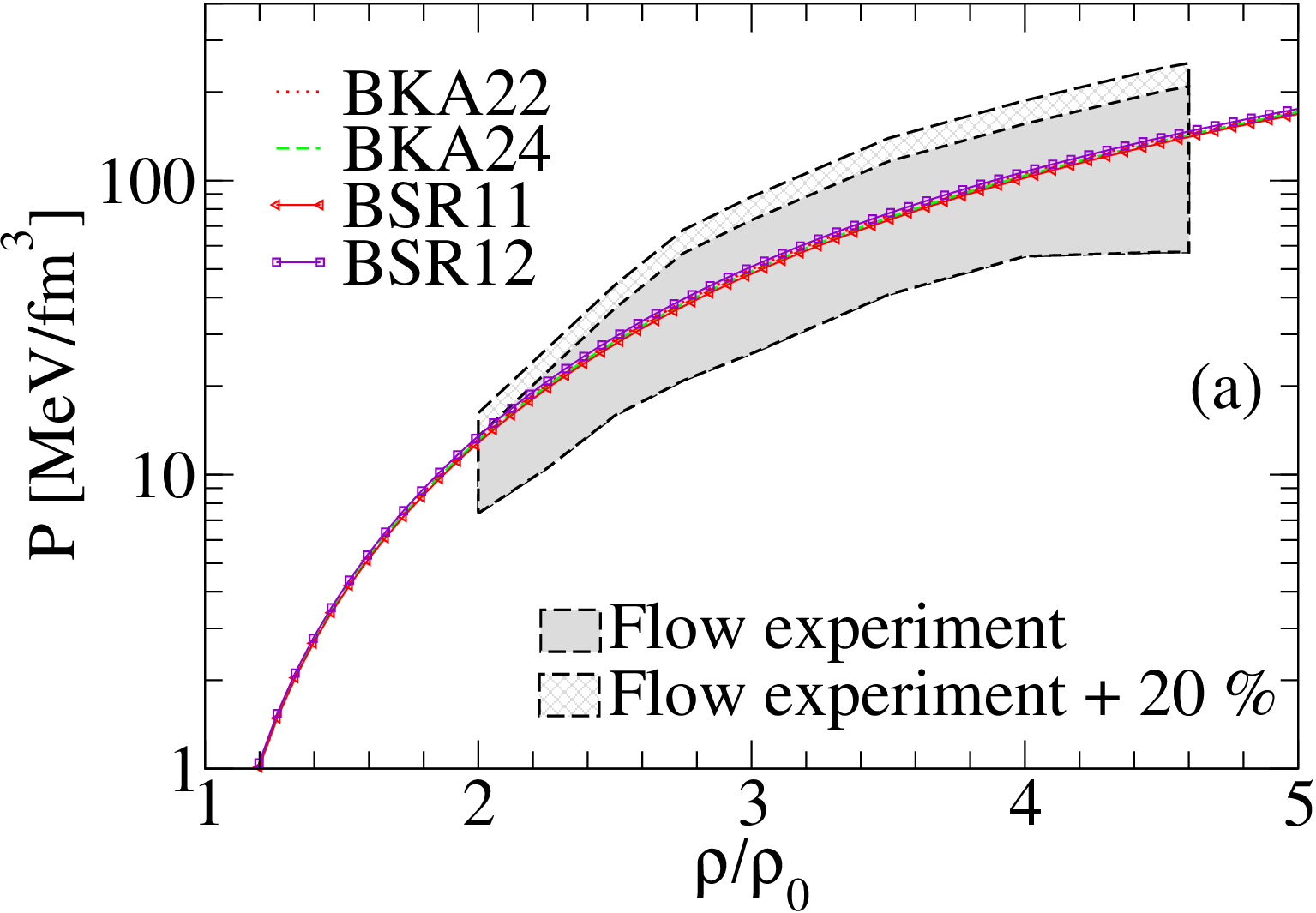}\\
\includegraphics[scale=0.24]{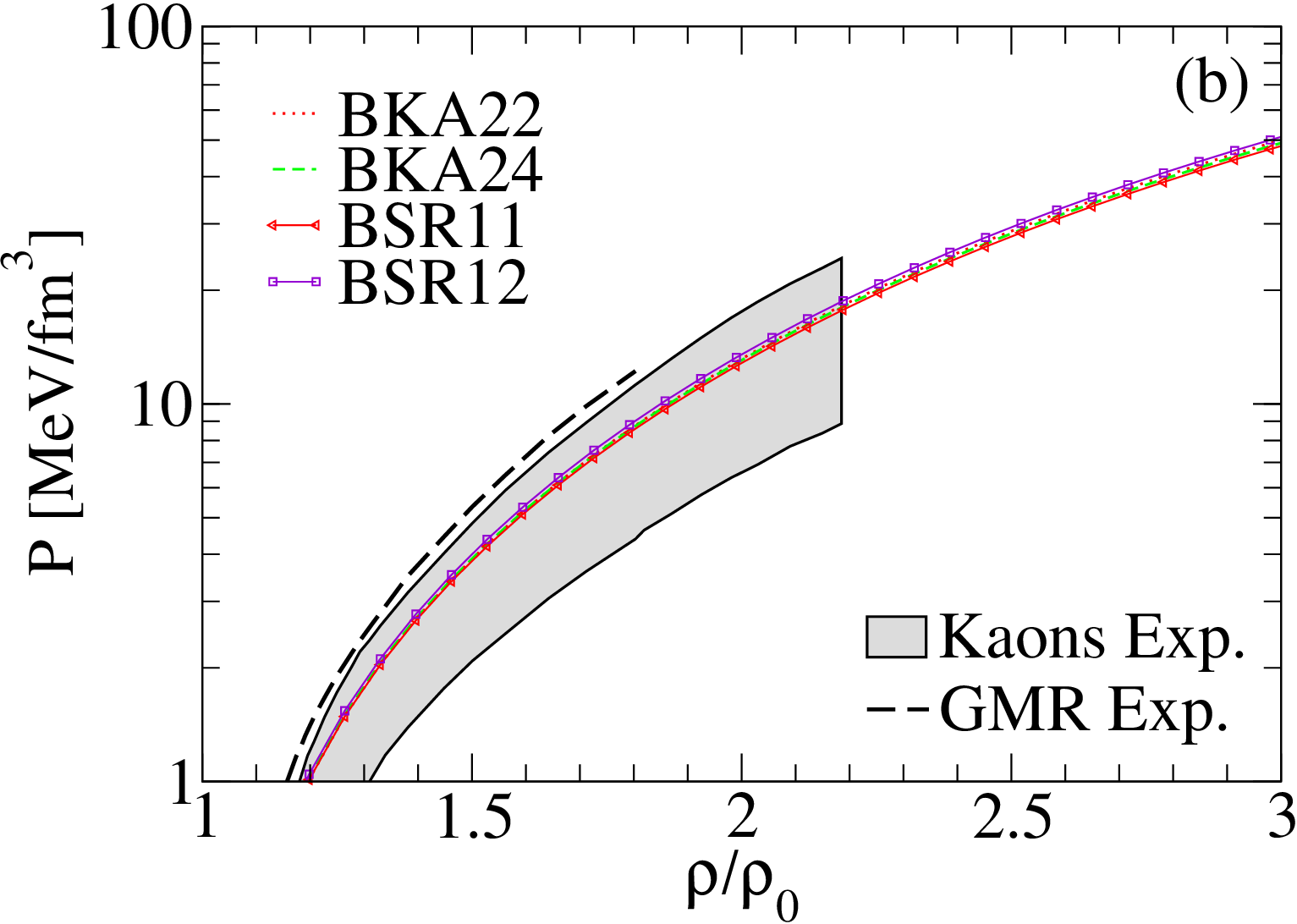}\\
\hspace{0.6cm}\includegraphics[scale=0.24]{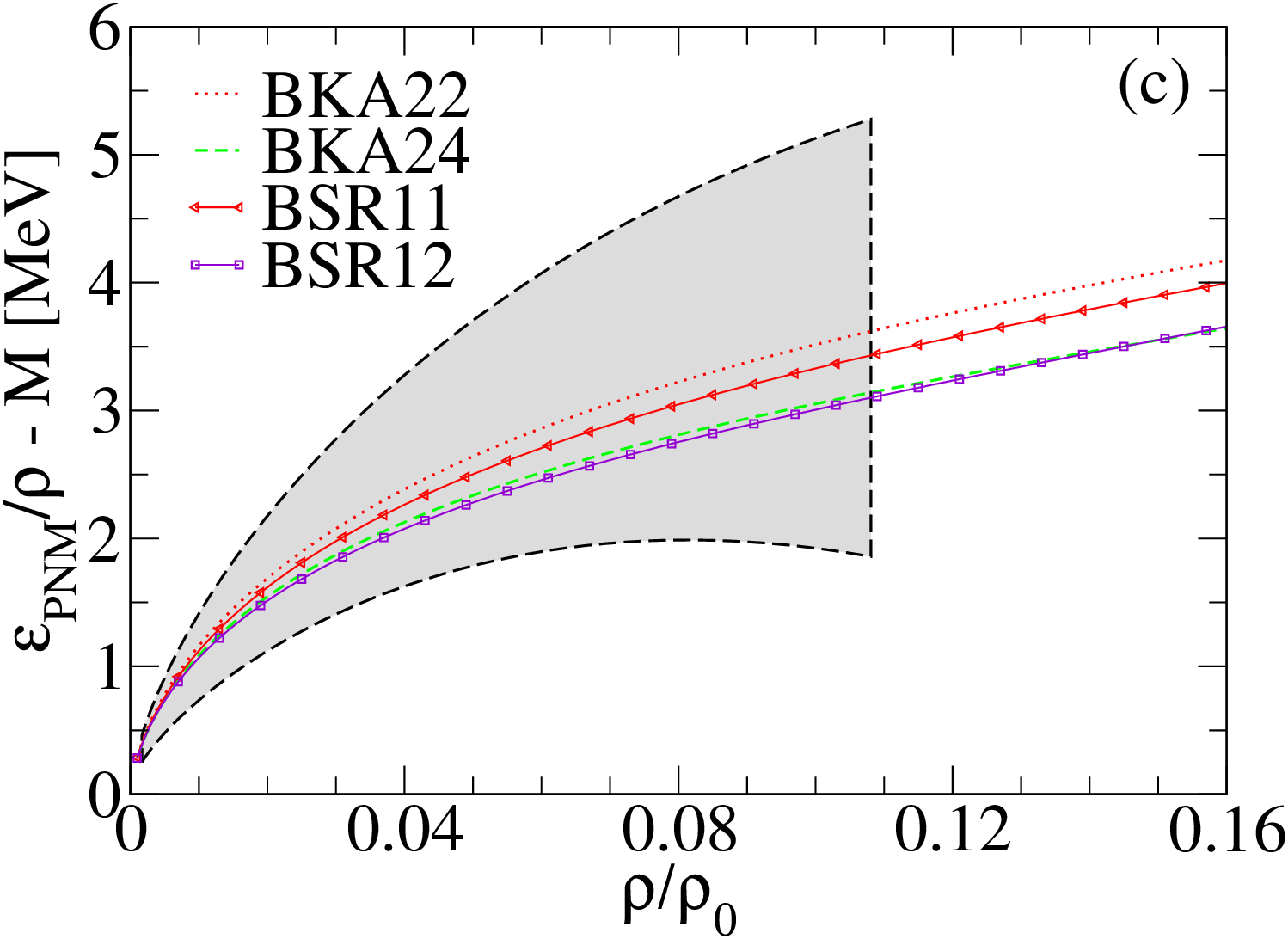}
\caption{(Color online) Density dependence of approved models in the a) SM3, b) SM4 and 
c) PNM1 constraints related to SET2a and SET2b. Shaded bands extracted from a) 
Ref.~\cite{SCI298-1592}, where flow experimental data is compared with 
results obtained for symmetric matter, b) Ref.~\cite{PPNP62-427}, where 
pressure in symmetric matter is compared with data extracted from kaon 
production and c) Ref.~\cite{PRC85-035201}, as explained in the text.}
\label{fig:set2}
\end{center}
\end{figure}

When considering the SET2b, the more stringent one, we found no significant changes
in the results. In such an analysis, {\it only} the {BSR12} model, among 
$263$, satisfies all SET2b simultaneously, and $22$ parameterizations satisfy $7$ of 
the $8$ constraints. Two such models fell outside the range of the unsatisfied 
constraint by less than $5\%$, namely, {BSR11} and {BKA22}, the same as in 
the SET2a analysis. The MIX3 constraint is again the one not satisfied for these two 
models. In summary, we have in total, {\it 3 models} consistent with the SET2b 
analysis.

It is worth noting that the difference between the two analyses regarding SET2a and
SET2b is the absence of the BKA24 model in the CRMF approved models of SET2b, in 
comparison to those approved in SET2a. The reason for this absence is the slope of 
the symmetry energy of the BKA24 model, $L_0=84.8$~MeV, clearly outside the 
MIX2b range, given by $L_0=55\pm 25$~MeV, by more than $5\%$.

As in the previous section, we present in the fifth column of Table~\ref{tab:set2} 
the range of the constraints of SET2, defined by the CRMF models. In the case of 
MIX3 and MIX4 constraints, the boundary models are different for both sets, namely, 
SET2a and SET2b. Therefore, we included the letter (a) or (b) in the values in order 
to identify the corresponding set. The density dependence of the approved models in 
both analyses is shown in Fig.~\ref{fig:set2}.

Also regarding our analysis of SET2a and SET2b, in Table~\ref{tab:napset2} we present 
the number of approved models in each constraint. 
\begin{table}[!hbt]
\begin{ruledtabular}
\caption{Number of approved models (among $263$) in each constraint of SET2.}
\centering
\begin{tabular}{lccccccccccc}
\multicolumn{11}{c}{SET2}\\
\hline
Constraints  & SM1   & SM3a  & SM3b   & SM4  & PNM1  & MIX1a & MIX1b & MIX2a  &
MIX2b  & MIX3  & MIX4 \\
Number of models     & $153$ & $129$ & $104$ & $153$ & $193$ & $174$ & $162$ &
$216$ & $72$ & $96$ & $65$ \\
\end{tabular}
\label{tab:napset2}
\end{ruledtabular}
\end{table}

We also provide in Table~\ref{tab:plus-minus} of Appendix~\ref{plusminus} the 
information whether each model of the $263$ ones is approved ($+$) or not ($-$) in 
the constraints of SET2. In Table~\ref{tab:percent} of Appendix~\ref{deviation}, we 
give the deviation as calculated in Eq.~(\ref{dev}) for the numerical constraints, 
as well as those obtained for the graphic constraints, all of them related to SET2.

In order to finish the analysis of SET2, we point out that our work concerns only 
nuclear matter. Due to the translational invariance, rotational symmetry and rotational 
invariance around the third axis in isospin space of the latter, we can constrain only 
some features of the Lagrangian and the approved models can not guarantee successful 
predictions for finite nuclei. From our analysis, one can see from 
Table~\ref{tab:plus-minus} that the models FSUGold, FSUZG03, DD-ME$\delta$ and 
\mbox{IU-FSU} are consistent with all constraints except MIX3, a constraint applicable in 
the region of saturation density (it applies only to SET2a for the \mbox{IU-FSU} model). 
These parameter sets provide quite good global fits to binding energies, charge radii, 
isotopic shifts and neutron skin thicknesses.

For the sake of completeness, we display in Table~\ref{tab:sat-cons}, several nuclear 
matter properties, at the saturation density, of the approved models in all sets 
analyzed here, namely, SET1, SET2a and SET2b.
\begin{table}[!htb]
\scriptsize
\begin{ruledtabular}
\caption{Nuclear matter properties at the saturation density $\rho_0$ of the RMF 
models consistent with the macroscopic constraints.}
\centering
\begin{tabular}{lccccccccccc}
Model & approved in SET& $\rho_0$ & $E_0$ & $K_0$ &  $m^*$ &
$K^{\prime}$ & $J$
& $L_0$  & $K_{\tau,\rm v}^0$ & $\frac{\mathcal{S}(\rho_0/2)}{J}$ & \\
&    & (fm$^{-3}$) & (MeV) & (MeV) & & (MeV) & (MeV) & (MeV) & (MeV) &  & 
\\ \hline
BKA22  & 2a and 2b & $0.147$ & $-15.91$ & $225.24$ & $0.61$ & $283.29$ & $33.17$ &
$78.79$ & $-382.46$ & $0.59$ \\ 
BKA24  & 2a & $0.147$ & $-15.95$ & $227.06$ & $0.60$ & $273.58$ & $34.19$ &
$84.80$ & $-421.55$ & $0.57$ \\ 
BSR11  & 2a and 2b & $0.147$ & $-16.08$ & $226.75$ & $0.61$ & $312.37$ & $33.69$ &
$78.78$ & $-388.86$ & $0.59$\\ 
BSR12  & 2a and 2b & $0.147$ & $-16.10$ & $232.35$ & $0.61$ & $290.31$ & $34.00$ &
$77.90$ & $-414.30$ & $0.59$\\ 
Z271v5  & 1 & $0.148$ & $-16.24$ & $271.00$ & $0.80$ & $733.59$ & $34.04$ &
$73.90$ & $-388.52$ & $0.58$ \\ 
Z271v6  & 1 & $0.148$ & $-16.24$ & $271.00$ & $0.80$ & $733.59$ & $33.80$ &
$70.94$ & $-388.36$ & $0.58$ \\ 
\end{tabular}
\label{tab:sat-cons}
\end{ruledtabular}
\end{table}

\subsection{Excluding MIX3 constraint}

The results presented in the previous section pointed to a small number of RMF models
consistent to all constraints simultaneously, namely, $4$ in the SET2a analysis, and $3$
for the SET2b one. However, these numbers change significantly if we simply discard the
MIX3 constraint from SET2a and SET2b. By excluding this constraint, the new results 
for the SET2a analysis become the following: {\it 25 models} are consistent with 
all~$7$ constraints. They are: {BKA20, BKA22, BKA24, BSR8, BSR9, BSR10, BSR11, 
BSR12, BSR15, BSR16, BSR17, BSR18, BSR19, \mbox{DD-ME$\delta$}, \mbox{DD-F}, 
DDH$\delta$, \mbox{FSU-III}, \mbox{FSU-IV}, FSUGold, FSUGold4, FSUGZ03, FSUGZ06, G2*, 
\mbox{IU-FSU},} and {TW99}. Moreover, $48$ models satisfy all but one of the 
constraints. In this group, $10$ models fell outside the range of the constraint by 
less than $5\%$. They are: {BSR20, FA3, Z271s2, Z271s3, Z271s4, Z271s5, Z271s6, 
Z271v4, Z271v5,} and {Z271v6}. By also including such models in the CRMF 
parameterizations, one has {\it 35~models} consistent with the SET2a analysis.

For the SET2b, we have {\it 22 models} consistent with all constraints. These are 
the same models approved in SET2a, except for the {{BKA24, \mbox{IU-FSU} and 
DDH$\delta$}} models. Moreover, $14$ models satisfy only $7$ constraints, and by 
applying the $5\%$ criterium, $8$ more models are approved. They are the same as in 
the corresponding case of SET2a, except for the {BSR20} and {FA3} models. Therefore, 
one has in the SET2b analysis a total of {\it 30 models} in the group of
approved models.

\subsection{Additional new constraints}

After the work presented in this manuscript has been completed, new constraints from
re-analysis of data on GMR energies became available~\cite{PRI-stone}, suggesting that the
SM1 constraint would span to somewhat higher values, $250$ - $315$~MeV (SM1-new) and MIX3
would be between $-620$ and $-370$~MeV (MIX3-new). We re-analysed the data introducing the
two additional constraints with the following results:
 
\subsubsection{Analysis considering MIX3-new in SET2}

\begin{itemize}

\item SET2a (SM1-new, SM3a, SM4, PNM1, MIX1a, MIX2a, MIX3-new and MIX4): $4$~models
approved, namely, {Z271v4, Z271v5, Z271v6}, and {FA3}. $24$~models not satisfying
only one constraint. Even with $5\%$ tolerance, no further model approved.
\item SET2b (SM1-new, SM3b, SM4, PNM1, MIX1b, MIX2b, MIX3-new and MIX4): $3$~models
approved, namely, {Z271v4, Z271v5}, and {Z271v6}. $12$~models not satisfying only
one constraint. Even with $5\%$ tolerance, no further model approved.

\end{itemize}

\subsubsection{Analysis discarding MIX3-new in SET2}

\begin{itemize}

\item SET2a (SM1-new, SM3a, SM4, PNM1, MIX1a, MIX2a, and MIX4): $9$~models approved,
namely, {Z271v4, Z271v5, Z271v6, Z271s2, Z271s3, Z271s4, Z271s5, Z271s6}, and
{FA3}. $45$~models not satisfying only one constraint. Even with $5\%$ tolerance, no
further model approved.

\item SET2b (SM1-new, SM3b, SM4, PNM1, MIX1b, MIX2b, and MIX4): $8$~models approved,
namely, {Z271v4, Z271v5, Z271v6, Z271s2, Z271s3, Z271s4, Z271s5}, and {Z271s6}.
$26$~models not satisfying only one constraint. Even with $5\%$ tolerance, no further
model approved.

\end{itemize}

\section{Summary and Conclusions}
\label{conclusions}

In this work we submitted $263$ parameterizations of the widely used relativistic 
mean-field hadronic models to three different sets of constraints. One of them 
(SET1) is composed exactly of the same constraints used in the extensive study of
Ref.~\cite{PRC85-035201} in which $240$ parameterizations of the nonrelativistic Skyrme
model were examined. The second one, named SET2, was divided in two other sets, in 
which updated constraints were taken into account and some of them removed from our 
analysis. The first set, SET2a, contains a broader version of the constraints and 
the second, SET2b, a more stringent one. In summary, all sets present constraints 
regarding information of symmetric nuclear matter (SM), pure neutron matter (PNM), 
and those in which these two frameworks are considered simultaneously (MIX), i. e., 
constraints derived from the symmetry energy at the saturation density. 

We have organized the RMF models in seven different groups regarding their Lagrangian 
density structure, namely, linear (type~$1$), $\sigma^3+\sigma^4$ (type~$2$), 
$\sigma^3+\sigma^4+\omega_0^4$ (type~$3$), $\sigma^3+\sigma^4+\omega_0^4+$ cross 
terms (type~$4$), density dependent (type~$5$), point-coupling (type~$6$), and delta 
meson (type~$7$) models. Their saturation properties are displayed in 
Table~\ref{tab:sat} of Appendix~\ref{apsatprop}.

The application of the SET1 constraints to the models, leads to the impressive result 
of only the Z271v5 and Z271v6 parameterizations being classified as 
consistent RMF (CRMF) models. This result is still more stringent than that found in 
Ref.~\cite{PRC85-035201} in which among $240$ Skyrme parameterizations, only $16$ 
were approved in the analysis regarding SET1. If we consider our analysis based on 
the constraints of SET2a, a more updated version in comparison with SET1, the 
results do not change significantly. In this case, BKA22, BKA24, BSR11 and BSR12 are 
the CRMF models, and, except for the BKA24 model, this is the same result when we 
use the SET2b, a more stringent one in comparison with the updated SET2a. The 
saturation properties of these models are summarized in Table~\ref{tab:sat-cons}. 
Also, the density dependence of the properties used to construct the graphic 
constraints are shown in Figs.~\ref{fig:set1} and~\ref{fig:set2}.

As an interesting feature concerning our analysis, we point out that in all sets of 
constraints, the approved models are of type~$4$, i. e., models presenting cross 
terms between the mesonic fields $\sigma$, $\omega_\mu$ and $\vec{\rho_\mu}$. In 
particular, these interactions include a   density dependence of the symmetry energy 
that goes beyond the almost linear behavior of models of the types 1, 2 and 3. 
Moreover, models BKA22 and BKA24 yield a neutron-skin thickness in the $^{208}$Pb 
nucleus, respectively, of 0.22 and 0.24 fm and were found to be consistent with  most 
of the constraints included in sets 2a and 2b \cite{PRC81-034323}. This means that 
models of type-4, those that include the cross term interactions, favor the 
reproduction of the expected behavior of the properties defined in each individual 
constraint. In order to satisfy the constraints used in this work, new 
parameterizations of the RMF models must take into account the density dependence of 
the symmetry energy through the inclusion of cross terms between the isoscalar 
mesons and the vector-isovector meson or other means.

We also highlight that such results change dramatically if we simply 
neglect from our analysis in SET2a and SET2b, the constraint regarding the volume 
part of the isospin incompressibility, $K_{\tau,\rm v}^0\approx 
K_\tau=-550\pm150$~MeV. By applying the $7$ remaining constraints of SET2a, we found 
the following $35$ approved models: BKA20, BKA22, BKA24, BSR8, BSR9, BSR10, BSR11, 
BSR12, BSR15, BSR16, BSR17, BSR18, BSR19, \mbox{DD-ME$\delta$}, \mbox{DD-F}, 
DDH$\delta$, \mbox{FSU-III}, \mbox{FSU-IV}, FSUGold, FSUGold4, FSUGZ03, FSUGZ06, G2*, 
\mbox{IU-FSU}, TW99. BSR20, FA3, Z271s2, Z271s3, Z271s4, Z271s5, Z271s6, Z271v4, 
Z271v5, and Z271v6. Notice that besides the cross term models, other types of 
models are consistent with the SET2a constraints, namely, two density dependent 
(\mbox{DD-F}, TW99), one point-coupling (FA3), and two delta meson models
(\mbox{DD-ME$\delta$}, DDH$\delta$). The common feature of these models is that 
they do not include the simple symmetry energy density dependence present in all type 
1, 2 and 3 models. The use of  isospin dependent properties in the fit procedure is 
essential to get a reasonable parametrization. For SET2b, the analysis leads to the 
same group of approved models excluding the BKA24, \mbox{IU-FSU}, DDH$\delta$, 
BSR20, and FA3 models, totaling $30$ consistent parameterizations. Notice that in 
this more restrictive set, one of the delta meson (DDH$\delta$) models and the 
point-coupling model are excluded from the CRMF models. However, the density 
dependent models \mbox{DD-F} and TW99 remain approved.

In the present work we have used the known physics of nuclear matter at densities 
around the saturation value to constrain relativistic models. However, these models 
have been extensively used to describe compact star constituents and macroscopic 
properties, such as radii and masses, and for this purpose have been extended to much 
higher densities. In this respect, the inclusion of other degrees of freedom, such as 
the lightest eight baryons, is energetically favored, but the onset of hyperons 
softens the equations of state and consequently reduces the maximum stellar masses 
\cite{compactstars}. Having in mind the description of two recently detected 
neutron stars with masses of the order of 2 $M_\odot$ \cite{demo,antoniadis}, 
appropriate equations of state with the inclusion of hyperons were shown to strongly 
depend on the choice of the models and also on the hyperon-meson coupling constants
\cite{schaffner,cp13,luiz2014}. After the comprehensive analyses performed in the
present work, we suggest that the models that passed all tests be used in further 
studies involving the inclusion of hyperons in astrophysical applications.  
         
We remark that there is still another class of relativistic models known as 
quark-meson-coupling models \cite{qmc}, in which baryons are described as a system 
of non-overlapping MIT bags interacting through intermediate mesons. In these models, 
quark degrees of freedom are explicitly taken into account and the couplings are 
determined at the quark level. These models will also be examined according to the 
constraints proposed in the present work in a future investigation.

Finally, we reinforce that our work represents a unique effort to qualify the 
current RMF and their usefulness for modeling of nuclear matter which has never 
been done before. At present it is impossible to critically  compare published 
results, obtained with different selection of RMF models because nuclear matter 
properties are model dependent. Consistent use of a narrowed selection of 
approved RMF model should allow to improve the situation. Furthermore, the 
failure of many RMF models to satisfy the most up-to-date constraints should 
stimulate search for missing physics, which, when included, should lead to 
improvement of the models and their predictive power.

As a final remark concerning the analysis performed in the present work (RMF models), 
and in the previous one (nonrelativistic Skyrme models~\cite{PRC85-035201}), we highlight 
that it would be desirable to establish a unique protocol to study the predictive power of 
mean-field models (both relativistic and nonrelativistic) to deal not only with nuclear 
matter but also with finite nuclei. Actually, Stone and Reinhard 
outlined a possible protocol for the Skyrme interaction in Sec. 5 of their 
review~\cite{PPNP58-587}, by indicating the finite nuclei constraints that should be used 
in order to select Skyrme parameterizations, such as charge r.m.s radius, spin-orbit 
splittings, neutron radii isotopic shifts, excitation properties and
others. Despite the fact that this 
procedure was developed for nonrelativistic models, it should also be applicable to RMF 
ones. Some of the entries should be updated, but the basic philosophy
remains the same. One of the 
positive outcomes of such a work would be to learn more about surface properties of 
finite nuclei. It is well known that the crucial difference between finite nuclei and 
nuclear matter is the presence (or absence) of the nuclear surface, which plays a very 
important role, but is not yet well understood. Comparison of the properties of nuclear matter 
around saturation density and those of finite nuclei could yield valuable information in 
this direction. A preliminary study involving these ideas was performed in 
Ref.~\cite{stevenson} for the Skyrme parametrizations selected in 
Ref.~\cite{PRC85-035201}. A more detailed study, also taking into account the selected RMF
models presented here, will be addressed in a future work.

\section*{ACKNOWLEDGMENTS}

This work was partially supported by CNPq (Brazil), CAPES (Brazil), FAPESP 
(Brazil) under projects 2009/00069-5, and FAPESC (Brazil) under project
2716/2012,TR 2012000344, by COMPETE/FEDER and FCT (Portugal) under Grant No.
PTDC/FIS/113292/2009 and by NEW COMPSTAR, a COST initiative. M. Dutra acknowledges the
hospitality at ITA/CTA and D.P.M. at the Universidad de Alicante, where parts of this work
were carried out. S. Typel is grateful for the support by the Helmholtz Association 
(HGF) through the Nuclear Astrophysics Virtual Institute (VH-VI-417). J.R.S. wishes to
thank Debora Menezes's group at Universidade Federal de Santa Catarina for its hospitality
during her stay, made possible by CNPq Processo (401593/2009-6), when this project was
initiated. O. L. acknowledges support from FAPESP.

\appendix
\section{Analytical expressions for infinite nuclear matter quantities}
\label{type4models}

In this appendix, we show the analytical expressions of some quantities 
at 
$y=1/2$ used to analyze the constraints of Sec.~\ref{resdis}, related to RMF 
models of types $4$
(nonlinear finite range), $5$ (density dependent) and $6$ (nonlinear
point-coupling). 
The quantities are obtained at zero temperature and generalized to any 
density.

\subsection{Finite range models}

The incompressibility is obtained as,
\begin{eqnarray}
K_{\rm NL} &=& 9\left(\frac{\partial P_{\rm NL}}{\partial\rho}\right)_{y=1/2} 
= 9\left[g_\omega\rho\frac{\partial\omega_0}{\partial\rho}
+ \frac{k_F^2}{3(k_F^2+{M^*}^2)^{1/2}} 
+ \frac{\rho M^*}{(k_F^2+{M^*}^2)^{1/2}}
\frac{\partial M^*}{\partial\rho}\right],
\end{eqnarray}
where
\begin{eqnarray} 
\frac{\partial M^*}{\partial\rho}=-g_\sigma\frac{\partial\sigma}{\partial\rho}.
\end{eqnarray}

The skewness coefficient is
\begin{eqnarray}
Q_{\rm NL} = 27\rho^3\frac{\partial^3(\mathcal{E}_{\rm
NL}/\rho)}{\partial\rho^3}\bigg|_{y=1/2}
= 27\rho^3\left[
\frac{1}{\rho}\frac{\partial^3\mathcal{E}_{\rm NL}}{\partial\rho^3}
-\frac{3}{\rho^2}\frac{\partial^2\mathcal{E}_{\rm NL}}{\partial\rho^2}
+\frac{6}{\rho^3}\frac{\partial\mathcal{E}_{\rm NL}}{\partial\rho}
-\frac{6\mathcal{E}_{\rm NL}}{\rho^4}\right]_{y=1/2},
\end{eqnarray}
with
\begin{eqnarray}
\frac{\partial\mathcal{E}_{\rm NL}}{\partial\rho} &=& (k_F^2+{M^*}^2)^{1/2} +
g_\omega\omega_0,
 \\
\frac{\partial^2\mathcal{E}_{\rm NL}}{\partial\rho^2} &=& 
g_\omega\frac{\partial\omega_0}{\partial\rho} 
+ \frac{1}{2E_F^*}\left(\frac{\pi^2}{k_F} 
+ 2M^*\frac{\partial M^*}{\partial\rho}\right)
\qquad\mbox{and}  \\
\frac{\partial^3\mathcal{E}_{\rm NL}}{\partial\rho^3} &=&
g_\omega\frac{\partial^2\omega_0}{\partial\rho^2}
-\frac{1}{4{E_F^*}^3}\left(\frac{\pi^2}{k_F} +
2M^*\frac{\partial M^*}{\partial\rho}\right)^2 \nonumber 
\\
&+& \frac{1}{2E_F^*}\left[-\frac{\pi^4}{2k_F^4}
+ 2\left(\frac{\partial M^*}{\partial\rho}\right)^2 
+ 2M^*\frac{\partial^2 M^*}{\partial\rho^2}\right].
\end{eqnarray}
The quantities $\mathcal{E}_{\rm NL}$, $\frac{\partial\mathcal{E}_{\rm
NL}}{\partial\rho}$, $\frac{\partial^2\mathcal{E}_{\rm NL}}{\partial\rho^2}$ and
$\frac{\partial^3\mathcal{E}_{\rm NL}}{\partial\rho^3}$ are evaluated at $y=1/2$.

The slope and curvature of symmetry energy are given, respectively, by
\begin{eqnarray}
L_{\rm NL} &=&
3\rho\left(\frac{\partial\mathcal{S}_{\rm NL}}{\partial\rho}\right)=\frac{k_F^2}{3E_F^* }
-\frac{k_F^4}{6{E_F^*}^3}\left(1+\frac{2M^*k_F}{\pi^2}\frac{\partial M^*}{\partial\rho}
\right)+\frac{3g_\rho^2}{8{m_\rho^*}^2}\rho-\frac{3g_\rho^2}{8{m_\rho^*}^4}\frac{\partial{
m_\rho^*}^2}{\partial\rho }\rho^2,\qquad
\end{eqnarray}
and
\begin{eqnarray}
K_{\rm sym}^{\rm NL} &=& 9\rho^2\left(\frac{\partial^2\mathcal{S}_{\rm
NL}}{\partial\rho^2}\right)
= 9\rho^2\left\lbrace  
-\frac{\pi^2}{12{E_F^*}^3k_F}\left(\frac{\pi^2}{k_F} 
+ 2M^*\frac{\partial M^*}{\partial\rho}\right) - \frac{\pi^4}{12E_F^*k_F^4}
-\frac{g_\rho^2}{4{m_\rho^*}^4}\frac{\partial{m_\rho^*}^2}{\partial\rho}
\right. \nonumber \\
&-& \left. \left[\frac{\pi^4}{24{E_F^*}^3k_F^2} -
\frac{k_F\pi^2}{8{E_F^*}^5}\left(\frac{\pi^2}{k_F} +
2M^*\frac{\partial M^*}{\partial\rho}\right)\right]
\left(1 + \frac{2M^*k_F}{\pi^2}\frac{\partial M^*}{\partial\rho}\right) 
+ \frac{g_\rho^2}{4{m_\rho^*}^6}\left(\frac{\partial{m_\rho^*}^2}
{\partial\rho}\right)^2\rho \right.\nonumber \\
&-& \left.
\frac{k_F\pi^2}{12{E_F^*}^3}\left[\frac{M^*}{k_F^2}\frac{\partial
M^*}{\partial\rho} + \frac{2k_F}{\pi^2}\left(\frac{\partial
M^*}{\partial\rho}\right)^2 + \frac{2k_F M^*}{\pi^2}\frac{\partial^2
M^*}{\partial\rho^2}\right]
-\frac{g_\rho^2}{8{m_\rho^*}^4}\frac{\partial^2{m_\rho^*}^2}{\partial\rho^2}\rho
\right\rbrace.
\end{eqnarray}

For symmetric nuclear matter, where $y=1/2$ and $\bar{\rho}_{0(3)}=0$, the density derivatives
of $\sigma$ and $\omega_0$ are given by
\begin{eqnarray}
\frac{\partial\sigma}{\partial\rho} =
\frac{a_1b_2+a_2b_3}{a_1b_1-a_3b_3}\quad\mbox{and}\quad 
\frac{\partial\omega_0}{\partial\rho} =
\frac{a_2b_1+a_3b_2}{a_1b_1-a_3b_3},
\end{eqnarray}
where
\begin{eqnarray}
a_1&=&m_\omega^2 + 3cg_\omega^4\omega_0^2 + g_\sigma
g_\omega^2\sigma(2\alpha_1+\alpha_1'g_\sigma\sigma), \\
a_2 &=& g_\omega, \\
a_3 &=& -2g_\sigma g_\omega^2\omega_0(\alpha_1+\alpha_1'g_\sigma\sigma), \\
b_1 &=& m_\sigma^2 + 2A\sigma + 3B\sigma^2 - g_\sigma^2g_\omega^2\omega_0^2\alpha_1' +
3g_\sigma^2\left(\frac{\rho_s}{M^*}-\frac{\rho}{E_F^*}\right), \\
b_2 &=& \frac{g_\sigma M^*}{E_F^*}\quad\mbox{and}\\
b_3 &=& -a_3.
\end{eqnarray}

For the sake of completeness, we also present the proton and neutron chemical potentials,
\begin{eqnarray}
\mu_p^{\rm NL} &=& \frac{\partial\mathcal{E}_{\rm NL}}{\partial\rho_p} =
(k_F^2+{M^*}^2)^{1/2} + g_\omega\omega_0 +\frac{g_\rho}{2}\bar{\rho}_{0(3)} \quad\mbox{and}
\\
\mu_n^{\rm NL} &=& \frac{\partial\mathcal{E}_{\rm NL}}{\partial\rho_n} =
(k_F^2+{M^*}^2)^{1/2} + g_\omega\omega_0 -
\frac{g_\rho}{2}\bar{\rho}_{0(3)},
\end{eqnarray}
both at any density and proton fraction $y$.

\subsection{Density dependent models}
The incompressibility is,
\begin{eqnarray}
K_{\rm DD} &=& 9\left(\frac{\partial P_{\rm DD}}{\partial\rho}\right)_{y=1/2} \nonumber\\
&=& 9\left[\rho\frac{\partial\Sigma_R}{\partial\rho} 
+ \omega_0\rho\frac{\partial\Gamma_\omega}{\partial\rho} 
+ \Gamma_\omega\rho\frac{\partial\omega_0}{ \partial\rho }
+ \frac{k_F^2}{3(k_F^2+{M^*}^2)^{1/2}} 
+ \frac{\rho M^*}{(k_F^2+{M^*}^2)^{1/2}}
\frac{\partial M^*}{\partial\rho}\right]\nonumber
\\
&=& 9\left[\rho\frac{\partial\Sigma_R}{\partial\rho} 
+ \frac{2\Gamma_\omega\rho^2}{m_\omega^2}\frac{\partial\Gamma_\omega}{\partial\rho} 
+ \frac{\Gamma_\omega^2\rho}{m_\omega^2}
+ \frac{k_F^2}{3(k_F^2+{M^*}^2)^{1/2}} 
+ \frac{\rho M^*}{(k_F^2+{M^*}^2)^{1/2}}
\frac{\partial M^*}{\partial\rho}\right],\qquad
\end{eqnarray}
with
\begin{eqnarray} 
\frac{\partial
M^*}{\partial\rho}&=&-\left(\Gamma_\sigma\frac{\partial\sigma}{\partial\rho}
+\sigma\frac{\partial\Gamma_\sigma}{\partial\rho}\right)\qquad\mbox{and}
\\
\frac{\partial\sigma}{\partial\rho}&=&\frac{\left[\rho_s-3\left(\frac{\rho_s}{M^*}
-\frac{\rho}{E_F^*}\right)\Gamma_\sigma\sigma\right]\frac{\partial\Gamma_\sigma}
{\partial\rho}+\frac{\Gamma_\sigma M^*}{E_F^*}}
{m_\sigma^2+3\left(\frac{\rho_s}{M^*}-\frac{\rho}{E_F^*}\right)\Gamma_\sigma^2},
\end{eqnarray}
observing that $\omega_0=\frac{\Gamma_\omega\rho}{m_\omega^2}$ and
$\sigma=\frac{\Gamma_\sigma\rho_s}{m_\sigma^2}$.

The skewness coefficient is
\begin{eqnarray}
Q_{\rm DD} = 27\rho^3\frac{\partial^3(\mathcal{E}_{\rm
DD}/\rho)}{\partial\rho^3}\bigg|_{y=1/2}
= 27\rho^3\left[
\frac{1}{\rho}\frac{\partial^3\mathcal{E}_{\rm DD}}{\partial\rho^3}
-\frac{3}{\rho^2}\frac{\partial^2\mathcal{E}_{\rm DD}}{\partial\rho^2}
+\frac{6}{\rho^3}\frac{\partial\mathcal{E}_{\rm DD}}{\partial\rho}
-\frac{6\mathcal{E}_{\rm DD}}{\rho^4}\right]_{y=1/2},\qquad
\end{eqnarray}
with
\begin{eqnarray}
\frac{\partial\mathcal{E}_{\rm DD}}{\partial\rho} &=& (k_F^2+{M^*}^2)^{1/2} +
\frac{\Gamma_\omega^2}{m_\omega^2}\rho + \bar{\Sigma}_{R},
 \\
\frac{\partial^2\mathcal{E}_{\rm DD}}{\partial\rho^2} &=& 
\frac{1}{2E_F^*}\left(\frac{\pi^2}{k_F} 
+ 2M^*\frac{\partial M^*}{\partial\rho}\right)
+\frac{\Gamma_\omega^2}{m_\omega^2}
+\frac{2\Gamma_\omega\rho}{m_\omega^2}\frac{\partial\Gamma_\omega}{\partial\rho}
+\frac{\partial\bar{\Sigma}_{R}}{\partial\rho}
\qquad\mbox{and}  
\\
\frac{\partial^3\mathcal{E}_{\rm DD}}{\partial\rho^3} &=&
-\frac{1}{4{E_F^*}^3}\left(\frac{\pi^2}{k_F} +
2M^*\frac{\partial M^*}{\partial\rho}\right)^2 
+ \frac{1}{2E_F^*}\left[-\frac{\pi^4}{2k_F^4}
+ 2\left(\frac{\partial M^*}{\partial\rho}\right)^2 
+ 2M^*\frac{\partial^2 M^*}{\partial\rho^2}\right]
\nonumber
\\
&+&\frac{2\Gamma_\omega\rho}{m_\omega^2}\frac{\partial^2\Gamma_\omega}{\partial\rho^2}
+\frac{2\rho}{m_\omega^2}\left(\frac{\partial\Gamma_\omega}{\partial\rho}\right)^2
+\frac{4\Gamma_\omega}{m_\omega^2}\frac{\partial\Gamma_\omega}{\partial\rho}
+\frac{\partial^2\bar{\Sigma}_{R}}{\partial\rho^2},
\end{eqnarray}
where
\begin{eqnarray}
\bar{\Sigma}_{R}&=&\Sigma_R(y=1/2)=\frac{\Gamma_\omega\rho^2}{m_\omega^2}
\frac{\partial\Gamma_\omega}{\partial\rho} -
\frac{\Gamma_\sigma\rho_s^2}{m_\sigma^2}\frac{\partial\Gamma_\sigma}{\partial\rho},
\\
\frac{\partial^2M^*}{\partial\rho^2}&=&
-\left(\Gamma_\sigma\frac{\partial^2\sigma}{\partial\rho^2}
+\sigma\frac{\partial^2\Gamma_\sigma}{\partial\rho^2}\right)
-2\frac{\partial\Gamma_\sigma}{\partial\rho}\frac{\partial\sigma}{\partial\rho},
\\
\frac{\partial^2\sigma}{\partial\rho^2}&=&\left[(\rho_s-\mathcal{J}
\Gamma_\sigma\sigma)\frac{\partial^2\Gamma_\sigma}{\partial\rho^2}
+\left(\frac{\partial\rho_s}{\partial\rho}-\Gamma_\sigma\sigma\frac{\partial\mathcal{J}}
{\partial\rho}-\mathcal{J}\sigma\frac{\partial\Gamma_\sigma}{\partial\rho}
-\mathcal{J}\Gamma_\sigma\frac{\partial\sigma}{\partial\rho}\right)\frac{
\partial\Gamma_\sigma}{\partial\rho}
+\frac{M^*}{E_F^*}\frac{\partial\Gamma_\sigma}{\partial\rho} \nonumber \right.
\\
&+&
\left.\frac{\Gamma_\sigma}{E_F^*}\frac{\partial M^*}{\partial\rho}
-\frac{\Gamma_\sigma M^*}{E_F^{*2}}\frac{\partial
E_F^*}{\partial\rho}\right](m_\sigma^2+\mathcal{J} \Gamma_\sigma^2)^{-1}\nonumber
\\
&-&\left[(\rho_s-\mathcal{J}\Gamma_\sigma\sigma)\frac{\partial\Gamma_\sigma}{\partial\rho}
+\frac{\Gamma_\sigma M^*}{E_F^*}\right]
\left[\frac{\partial\mathcal{J}}{\partial\rho}\Gamma_\sigma^2+2\mathcal{J}
\Gamma_\sigma\frac{\partial\Gamma_\sigma}{\partial\rho}\right](m_\sigma^2+\mathcal{J}
\Gamma_\sigma^2)^{-2}, 
\\
\mathcal{J}&=&3\left(\frac{\rho_s}{M^*} -\frac{\rho}{E_F^*}\right), 
\\
\frac{\partial E_F^*}{\partial\rho}&=&\frac{\pi^2}{2E_F^*k_F}
\left(1+\frac{2M^*k_F}{\pi^2}\frac{\partial M^*}{\partial\rho}\right),
\\
\frac{\partial^2\bar{\Sigma}_{R}}{\partial\rho^2}&=&\frac{\Gamma_\omega\rho^2}{
m_\omega^2}
\frac{\partial^3\Gamma_\omega}{\partial\rho^3}+\frac{4\Gamma_\omega\rho}{m_\omega^2}\frac{
\partial^2\Gamma_\omega}{\partial\rho^2}+\frac{3\rho^2}{m_\omega^2}
\frac{\partial\Gamma_\omega}{\partial\rho}\frac{\partial^2\Gamma_\omega}{\partial\rho^2}
+\frac{4\rho}{m_\omega^2}\left(\frac{\partial\Gamma_\omega}{\partial\rho}\right)^2
+\frac{2\Gamma_\omega}{m_\omega^2}\frac{\partial\Gamma_\omega}{\partial\rho}\nonumber
\\
&-&\frac{\Gamma_\sigma\rho_s^2}{m_\sigma^2}\frac{\partial^3\Gamma_\sigma}{\partial\rho^3}
-\frac{4\Gamma_\sigma\rho_s}{m_\sigma^2}\frac{\partial\rho_s}{\partial\rho}
\frac{\partial^2\Gamma_\sigma}{\partial\rho^2}
-\frac{3\rho_s^2}{m_\sigma^2}\frac{\partial\Gamma_\sigma}{\partial\rho}
\frac{\partial^2\Gamma_\sigma}{\partial\rho^2}
-\frac{4\rho_s}{m_\sigma^2}\frac{\partial\rho_s}{\partial\rho}\left(
\frac{\partial\Gamma_\sigma}{\partial\rho}\right)^2 \nonumber 
\\
&-&\frac{2\Gamma_\sigma}{m_\sigma^2}\left(\frac{\partial\rho_s}{\partial\rho}\right)^2
\frac{\partial\Gamma_\sigma}{\partial\rho}
-\frac{2\Gamma_\sigma\rho_s}{m_\sigma^2}\frac{\partial^2\rho_s}{\partial\rho^2}
\frac{\partial\Gamma_\sigma}{\partial\rho},
\\
\frac{\partial\rho_s}{\partial\rho}&=&\frac{M^*}{E_F^*}
-\mathcal{J}\left(\sigma\frac{\partial\Gamma_\sigma}{\partial\rho}
+\Gamma_\sigma\frac{\partial\sigma}{\partial\rho}\right),
\end{eqnarray}
and
\begin{eqnarray}
\frac{\partial^2\rho_s}{\partial\rho^2}&=&\frac{1}{E_F^*}\frac{\partial M^*}{\partial\rho}
-\frac{M^*}{E_F^{*2}}\frac{\partial E_F^*}{\partial\rho}
-\mathcal{J}\left(\sigma\frac{\partial^2\Gamma_\sigma}{\partial\rho^2}
+2\frac{\partial\Gamma_\sigma}{\partial\rho}\frac{\partial\sigma}{\partial\rho}
+\Gamma_\sigma\frac{\partial^2\sigma}{\partial\rho^2}\right)\nonumber
\\
&-&\frac{\partial\mathcal{J}}{\partial\rho}
\left(\sigma\frac{\partial\Gamma_\sigma}{\partial\rho}
+\Gamma_\sigma\frac{\partial\sigma}{\partial\rho}\right).
\end{eqnarray}

The slope and curvature of $\mathcal{S}_{\rm DD}$ are
\begin{eqnarray}
L_{\rm DD} &=&
3\rho\left(\frac{\partial\mathcal{S}_{\rm DD}}{\partial\rho}\right)=\frac{k_F^2}{3E_F^* }
-\frac{k_F^4}{6{E_F^*}^3}\left(1+\frac{2M^*k_F}{\pi^2}\frac{\partial M^*}{\partial\rho}
\right)+\frac{3\Gamma_\rho^2}{8{m_\rho}^2}\rho 
+ \frac{3\Gamma_\rho}{4m_\rho^2}\frac{\partial\Gamma_\rho}{\partial\rho}\rho^2,\qquad
\end{eqnarray}
and
\begin{eqnarray}
K_{\rm sym}^{\rm DD} &=& 9\rho^2\left(\frac{\partial^2\mathcal{S}_{\rm
DD}}{\partial\rho^2}\right)
= 9\rho^2\left\lbrace  
-\frac{\pi^2}{12{E_F^*}^3k_F}\left(\frac{\pi^2}{k_F} 
+ 2M^*\frac{\partial M^*}{\partial\rho}\right) - \frac{\pi^4}{12E_F^*k_F^4}
+\frac{\Gamma_\rho}{2m_\rho^2}\frac{\partial\Gamma_\rho}{\partial\rho}
\right. \nonumber \\
&-& \left. \left[\frac{\pi^4}{24{E_F^*}^3k_F^2} -
\frac{k_F\pi^2}{8{E_F^*}^5}\left(\frac{\pi^2}{k_F} +
2M^*\frac{\partial M^*}{\partial\rho}\right)\right]
\left(1 + \frac{2M^*k_F}{\pi^2}\frac{\partial M^*}{\partial\rho}\right) 
+ \frac{\rho}{4m_\rho^2}\left(\frac{\partial\Gamma_\rho}
{\partial\rho}\right)^2 \right.\nonumber \\
&-& \left.
\frac{k_F\pi^2}{12{E_F^*}^3}\left[\frac{M^*}{k_F^2}\frac{\partial
M^*}{\partial\rho} + \frac{2k_F}{\pi^2}\left(\frac{\partial
M^*}{\partial\rho}\right)^2 + \frac{2k_F M^*}{\pi^2}\frac{\partial^2
M^*}{\partial\rho^2}\right]
+\frac{\Gamma_\rho\rho}{4m_\rho^2}\frac{\partial^2\Gamma_\rho}{\partial\rho^2}
\right\rbrace.
\end{eqnarray}

The chemical potentials for any proton fraction are,
\begin{eqnarray}
\mu_p^{\rm DD} &=& \frac{\partial\mathcal{E}_{\rm DD}}{\partial\rho_p} =
(k_F^2+{M^*}^2)^{1/2} + \Gamma_\omega\omega_0 +\frac{\Gamma_\rho}{2}\bar{\rho}_{0(3)}
+\Sigma_R
\quad\mbox{and}
\\
\mu_n^{\rm DD} &=& \frac{\partial\mathcal{E}_{\rm DD}}{\partial\rho_n} =
(k_F^2+{M^*}^2)^{1/2} + \Gamma_\omega\omega_0 -\frac{\Gamma_\rho}{2}\bar{\rho}_{0(3)}
+ \Sigma_R.
\end{eqnarray}

\subsection{Point-coupling models}

For NLPC models, the incompressibility reads
\begin{eqnarray}
K_{\rm NLPC} &=& 9\left(\frac{\partial P_{\rm
NLPC}}{\partial\rho}\right)_{y=1/2}
= 9\left[\alpha_{\mbox{\tiny V}}\rho + 3\gamma_{\mbox{\tiny V}}\rho^3 
+ \frac{k_F^2}{3(k_F^2+{M^*}^2)^{1/2}} 
+ \frac{\rho M^*}{(k_F^2+{M^*}^2)^{1/2}}\frac{\partial M^*}{\partial\rho}\right.
\nonumber\\
&+& \left. 2(\eta_1+\eta_2\rho_s)\rho_s\rho
+2(\eta_1+2\eta_2\rho_s)\rho^2\frac{\partial\rho_s}{\partial\rho}\right],
\end{eqnarray}
where
\begin{eqnarray} 
\frac{\partial M^*}{\partial\rho}&=&\frac{2(\eta_1+2\eta_2\rho_s)\rho
+\frac{M^*}{E_F^*}(\alpha_s+2\beta_s\rho_s + 3\gamma_s\rho_s^2+2\eta_2\rho^2)}
{1-3(\alpha_s+2\beta_s\rho_s+
3\gamma_s\rho_s^2+2\eta_2\rho^2)\left(\frac{\rho_s}{M^*}-\frac{\rho}{E_F^*}\right)},
\\
\frac{\partial\rho_s}{\partial\rho}&=&\frac{M^*}{(k_F^2+M^{*2})^{1/2}}
+3\left(\frac{\rho_s}{M^*}-\frac{\rho}{E_F^*}\right)\frac{\partial M^*}{\partial\rho}.
\end{eqnarray}

The skewness coefficient is
\begin{eqnarray}
Q_{\rm NLPC}&=& 27\rho^3\frac{\partial^3(\mathcal{E}_{\rm
NLPC}/\rho)}{\partial\rho^3}\bigg|_{y=1/2}\nonumber\\
&=& 27\rho^3\left[
\frac{1}{\rho}\frac{\partial^3\mathcal{E}_{\rm NLPC}}{\partial\rho^3}
-\frac{3}{\rho^2}\frac{\partial^2\mathcal{E}_{\rm NLPC}}{\partial\rho^2}
+\frac{6}{\rho^3}\frac{\partial\mathcal{E}_{\rm NLPC}}{\partial\rho}
-\frac{6\mathcal{E}_{\rm NLPC}}{\rho^4}\right]_{y=1/2},\qquad
\end{eqnarray}
with
\begin{eqnarray}
\frac{\partial\mathcal{E}_{\rm NLPC}}{\partial\rho} &=& (k_F^2+{M^*}^2)^{1/2} +
\alpha_{\mbox{\tiny V}}\rho + \gamma_{\mbox{\tiny V}}\rho^3
+2(\eta_1+\eta_2\rho_s)\rho_s\rho,
 \\
\frac{\partial^2\mathcal{E}_{\rm NLPC}}{\partial\rho^2} &=& 
\frac{1}{2E_F^*}\left(\frac{\pi^2}{k_F} 
+ 2M^*\frac{\partial M^*}{\partial\rho}\right)
+\alpha_{\mbox{\tiny V}} + 3\gamma_{\mbox{\tiny V}}\rho^2
+2(\eta_1+2\eta_2\rho_s)\rho\frac{\partial\rho_s}{\partial\rho}
\nonumber\\
&+&2(\eta_1+\eta_2\rho_s)\rho_s, \qquad\mbox{and}  
\\
\frac{\partial^3\mathcal{E}_{\rm NLPC}}{\partial\rho^3} &=&
-\frac{1}{4{E_F^*}^3}\left(\frac{\pi^2}{k_F} +
2M^*\frac{\partial M^*}{\partial\rho}\right)^2 
+ \frac{1}{2E_F^*}\left[-\frac{\pi^4}{2k_F^4}
+ 2\left(\frac{\partial M^*}{\partial\rho}\right)^2 
+ 2M^*\frac{\partial^2 M^*}{\partial\rho^2}\right]
\nonumber
\\
&+& 6\gamma_{\mbox{\tiny V}}\rho + 4(\eta_1+2\eta_2\rho_s)\frac{\partial\rho_s}{\partial\rho}
+4\eta_2\rho\left(\frac{\partial\rho_s}{\partial\rho}\right)^2
+2(\eta_1+2\eta_2\rho_s)\rho\frac{\partial^2\rho_s}{\partial\rho^2}. 
\end{eqnarray}

The slope and curvature of $\mathcal{S}_{\rm NLPC}$ are
\begin{eqnarray}
L_{\rm NLPC} &=&
3\rho\left(\frac{\partial\mathcal{S}_{\rm NLPC}}{\partial\rho}\right)\nonumber\\
&=&\frac{k_F^2}{3E_F^*}
-\frac{k_F^4}{6{E_F^*}^3}\left(1+\frac{2M^*k_F}{\pi^2}\frac{\partial M^*}{\partial\rho}
\right)
+\frac{3}{2}\alpha_{\mbox{\tiny TV}}\rho+3\eta_3\rho_s\rho+3\eta_3\rho^2
\frac{\partial\rho_s}{\partial\rho},
\end{eqnarray}
and
\begin{eqnarray}
K_{\rm sym}^{\rm NLPC} &=& 9\rho^2\left(\frac{\partial^2\mathcal{S}_{\rm
NLPC}}{\partial\rho^2}\right)
= 9\rho^2\left\lbrace  
-\frac{\pi^2}{12{E_F^*}^3k_F}\left(\frac{\pi^2}{k_F} 
+ 2M^*\frac{\partial M^*}{\partial\rho}\right) - \frac{\pi^4}{12E_F^*k_F^4}
+2\eta_3\frac{\partial\rho_s}{\partial\rho}
\right. \nonumber \\
&-& \left. \left[\frac{\pi^4}{24{E_F^*}^3k_F^2} -
\frac{k_F\pi^2}{8{E_F^*}^5}\left(\frac{\pi^2}{k_F} +
2M^*\frac{\partial M^*}{\partial\rho}\right)\right]
\left(1 + \frac{2M^*k_F}{\pi^2}\frac{\partial M^*}{\partial\rho}\right) 
+ \eta_3\rho\frac{\partial^2\rho_s}{\partial\rho^2}\right.\nonumber \\
&-& \left.
\frac{k_F\pi^2}{12{E_F^*}^3}\left[\frac{M^*}{k_F^2}\frac{\partial
M^*}{\partial\rho} + \frac{2k_F}{\pi^2}\left(\frac{\partial
M^*}{\partial\rho}\right)^2 + \frac{2k_F M^*}{\pi^2}\frac{\partial^2
M^*}{\partial\rho^2}\right]\right\rbrace.
\end{eqnarray}

The chemical potentials for any proton fraction are,
\begin{eqnarray}
\mu_p^{\rm NLPC} &=& \frac{\partial\mathcal{E}_{\rm NLPC}}{\partial\rho_p} \nonumber\\
&=&(k_F^2+{M^*}^2)^{1/2} + \alpha_{\mbox{\tiny V}}\rho + \gamma_{\mbox{\tiny V}}\rho^3
+ \alpha_{\mbox{\tiny TV}}\rho_3 + \gamma_{\mbox{\tiny TV}}\rho_3^3 + 2\eta_1\rho_s\rho
+ 2\eta_2\rho_s^2\rho + 2\eta_3\rho_s\rho_3,\qquad
\\
\mu_n^{\rm NLPC} &=& \frac{\partial\mathcal{E}_{\rm NLPC}}{\partial\rho_n} \nonumber\\
&=&(k_F^2+{M^*}^2)^{1/2} + \alpha_{\mbox{\tiny V}}\rho + \gamma_{\mbox{\tiny V}}\rho^3
- \alpha_{\mbox{\tiny TV}}\rho_3 - \gamma_{\mbox{\tiny TV}}\rho_3^3 + 2\eta_1\rho_s\rho
+ 2\eta_2\rho_s^2\rho - 2\eta_3\rho_s\rho_3.\qquad
\end{eqnarray}

\section{Saturation Properties}
\label{apsatprop}

\begingroup
\centering
\squeezetable
\begin{longtable}{lccccrrrrrr}
\caption{Saturation properties of all RMF models used in this work. All
quantities are given in MeV, except for the dimensionless effective 
mass, $m^* =
M^*/M$, and the saturation density $\rho_0$, given in fm$^{-3}$. Here $K^\prime=-Q_0.$}
\label{tab:sat} \\
\hline\hline
\multicolumn{1}{l}{Model} & \multicolumn{1}{c}{$\rho_0$}
& \multicolumn{1}{c}{$E_0$} & \multicolumn{1}{c}{$K_0$}
& \multicolumn{1}{c}{$m^*$} & \multicolumn{1}{c}{$K'$}
& \multicolumn{1}{c}{$J$}  & \multicolumn{1}{c}{$L_0$} 
& \multicolumn{1}{c}{$K_{\rm sym}^0$} & \multicolumn{1}{c}{$Q_{\rm sym}^0$} 
& \multicolumn{1}{c}{$K_{\tau,\rm v}^0$} \\
\hline 
\endfirsthead
\multicolumn{11}{c}
{\tablename\ \thetable{} -- continued from previous page} \\
\hline\hline
\multicolumn{1}{l}{Model} & \multicolumn{1}{c}{$\rho_0$}
& \multicolumn{1}{c}{$E_0$} & \multicolumn{1}{c}{$K_0$}
& \multicolumn{1}{c}{$m^*$} & \multicolumn{1}{c}{$K'$}
& \multicolumn{1}{c}{$J$}  & \multicolumn{1}{c}{$L_0$} 
& \multicolumn{1}{c}{$K_{\rm sym}^0$} & \multicolumn{1}{c}{$Q_{\rm sym}^0$} 
& \multicolumn{1}{c}{$K_{\tau,\rm v}^0$} \\
\hline
\endhead
\hline
\multicolumn{11}{r}{{Continued on next page}} \\ 
\hline
\endfoot
\hline\hline
\endlastfoot
\multicolumn{11}{c}{linear finite range models (type 1)} \\
\hline
H1 \cite{EL48-505} & ~$0.148$~ & ~~$-15.75$~~ & ~~$546.81$~~ & $0.54$ & 
~~$-2152.62$ & ~~~$25.93$ & ~~~$88.38$ & ~~$93.56$ & ~~$-538.40$ &
~~$-784.64$\\ 
L1 \cite{RPP52-439} & $0.177$ & $-18.52$ & $625.56$ & $0.53$ & $-2229.20$ &
$21.68$ & $75.64$ & $81.88$ & $-786.15$ & $-641.56$ \\ 
L2 \cite{RPP52-439} & $0.142$ & $-16.78$ & $578.49$ & $0.53$ & $-2323.13$ &
$19.07$ & $68.81$ & $97.43$ & $-605.30$ & $-591.73$ \\ 
L3 \cite{RPP52-439} & $0.134$ & $-18.24$ & $624.53$ & $0.52$ & $-2568.70$ &
$18.86$ & $69.52$ & $102.17$ & $-709.63$ & $-600.92$ \\ 
LBF \cite{PRI-stone2} & $0.152$ & $-17.01$ & $584.18$ & $0.53$ & $-2261.30$ &
$46.92$ & $151.91$ & $92.59$ & $-640.84$ & $-1406.85$ \\ 
LHS \cite{RPP52-439} & $0.148$ & $-15.79$ & $548.10$ & $0.54$ & $-2157.94$ &
$34.99$ & $115.60$ & $93.62$ & $-541.51$ & $-1055.08$ \\ 
LW \cite{RPP52-439} & $0.194$ & $-15.75$ & $545.55$ & $0.56$ & $-1877.70$ &
$22.11$ & $74.44$ & $74.96$ & $-604.39$ & $-627.91$ \\
\end{longtable}
\pagebreak
\endgroup

\newpage
\begin{landscape}
\section{Plus-Minus Table}
\label{plusminus}
\begingroup
\pagestyle{headings}
\centering
\squeezetable
\begin{longtable}{lccccccccccclccccccccccc}
\caption{Status of each RMF model: approved ($+$) or not ($-$) under SET2a
and SET2b constraints. SET2a (SET2b): SM1, SM3a (SM3b), SM4, PNM1, 
MIX1a (MIX1b), MIX2a (MIX2b), MIX3 and MIX4.}
\label{tab:plus-minus} \\
\hline\hline
\multicolumn{1}{l}{Model}   & \multicolumn{1}{c}{SM1}
& \multicolumn{1}{c}{SM3a}   & \multicolumn{1}{c}{SM3b}
& \multicolumn{1}{c}{SM4}   & \multicolumn{1}{c}{PNM1}
& \multicolumn{1}{c}{MIX1a} & \multicolumn{1}{c}{MIX1b}
& \multicolumn{1}{c}{MIX2a} & \multicolumn{1}{c}{MIX2b} 
& \multicolumn{1}{c}{MIX3}  & \multicolumn{1}{c}{MIX4}
&\multicolumn{1}{l}{Model}   & \multicolumn{1}{c}{SM1}
& \multicolumn{1}{c}{SM3a}   & \multicolumn{1}{c}{SM3b}
& \multicolumn{1}{c}{SM4}   & \multicolumn{1}{c}{PNM1}
& \multicolumn{1}{c}{MIX1a} & \multicolumn{1}{c}{MIX1b}
& \multicolumn{1}{c}{MIX2a} & \multicolumn{1}{c}{MIX2b} 
& \multicolumn{1}{c}{MIX3}  & \multicolumn{1}{c}{MIX4} \\
\hline 
\endfirsthead
\multicolumn{24}{c}%
{\tablename\ \thetable{} -- continued from previous page} \\
\hline\hline
\multicolumn{1}{l}{Model}   & \multicolumn{1}{c}{SM1}
& \multicolumn{1}{c}{SM3a}   & \multicolumn{1}{c}{SM3b}
& \multicolumn{1}{c}{SM4}   & \multicolumn{1}{c}{PNM1}
& \multicolumn{1}{c}{MIX1a} & \multicolumn{1}{c}{MIX1b}
& \multicolumn{1}{c}{MIX2a} & \multicolumn{1}{c}{MIX2b} 
& \multicolumn{1}{c}{MIX3}  & \multicolumn{1}{c}{MIX4}
&\multicolumn{1}{l}{Model}   & \multicolumn{1}{c}{SM1}
& \multicolumn{1}{c}{SM3a}   & \multicolumn{1}{c}{SM3b}
& \multicolumn{1}{c}{SM4}   & \multicolumn{1}{c}{PNM1}
& \multicolumn{1}{c}{MIX1a} & \multicolumn{1}{c}{MIX1b}
& \multicolumn{1}{c}{MIX2a} & \multicolumn{1}{c}{MIX2b} 
& \multicolumn{1}{c}{MIX3}  & \multicolumn{1}{c}{MIX4} \\
\hline 
\endhead
\hline
\multicolumn{24}{l}{{Continued on next page}} \\ 
\hline
\endfoot
\hline\hline
\endlastfoot
\multicolumn{24}{c}{linear finite range models} \\
\hline
H1 & $-$ & $-$ & $-$ & $-$ & $+$ & $+$ & $-$ & $+$ & $-$ & $-$ & $-$ & LZ & $-$ & $-$ & $-$ & $-$ & 
$+$ & $-$ & $-$ & $-$ & $-$ & $-$ & $-$ \\ 
L1 & $-$ & $-$ & $-$ & $-$ & $+$ & $-$ & $-$ & $+$ & $+$ & $+$ & $-$ & RMF201 & $-$ & $-$ & $-$ & 
$-$ & $+$ & $+$ & $+$ & $+$ & $-$ & $-$ & $-$ \\ 
L2 & $-$ & $-$ & $-$ & $-$ & $-$ & $-$ & $-$ & $+$ & $+$ & $+$ & $-$ & RMF202 & $-$ & $-$ & $-$ & 
$-$ & $+$ & $+$ & $+$ & $+$ & $-$ & $-$ & $-$ \\ 
L3 & $-$ & $-$ & $-$ & $-$ & $-$ & $-$ & $-$ & $+$ & $+$ & $+$ & $-$ & RMF203 & $-$ & $-$ & $-$ & 
$-$ & $+$ & $+$ & $+$ & $+$ & $-$ & $-$ & $-$ \\ 
LBF & $-$ & $-$ & $-$ & $-$ & $+$ & $-$ & $-$ & $-$ & $-$ & $-$ & $-$ & RMF204 & $-$ & $-$ & $-$ & 
$-$ & $+$ & $+$ & $+$ & $+$ & $-$ & $-$ & $-$ \\ 
LHS & $-$ & $-$ & $-$ & $-$ & $+$ & $+$ & $+$ & $-$ & $-$ & $-$ & $-$ & RMF205 & $-$ & $-$ & $-$ & 
$-$ & $+$ & $+$ & $+$ & $+$ & $-$ & $-$ & $-$ \\ 
LW & $-$ & $-$ & $-$ & $-$ & $+$ & $-$ & $-$ & $+$ & $+$ &  $+$ & $-$ & RMF206 & $-$ & $-$ & $-$ & 
$-$ & $+$ & $+$ & $+$ & $+$ & $-$ & $-$ & $-$ \\
\end{longtable}
\pagebreak
\endgroup
\end{landscape}

\newpage
\section{Deviation Table}
\label{deviation}
\begingroup
\centering
\squeezetable
\begin{longtable}{lrcccccrrrrr}
\caption{List of the RMF parameterizations and their compliance with
macroscopic constraints used in this work. For SM1 and MIX1-4 we give the
standard deviation; the constraint is satisfied if the standard deviation is
less or equal to one. For SM3, SM4, 
and PNM1 we give the fraction of the full
density range in \% in which the constraint is not satisfied. A letter L 
(H)
indicates that the RMF prediction starts to fail at the 
beginning (end) of
the density range. No letter indication means that the model fails in the
middle of the density range. For models that fail in $100\%$ of the 
range, the letter U (D)
indicates that its curve is above (below) the band defined by the constraint.
For more details see the text.}
\label{tab:percent} \\
\hline\hline
\multicolumn{1}{l}{Model}         & \multicolumn{1}{r}{SM1}
& \multicolumn{1}{c}{SM3a($\%$)}  & \multicolumn{1}{c}{SM3b($\%$)}
& \multicolumn{1}{c}{SM4($\%$)}   & \multicolumn{1}{c}{PNM1($\%$)}
& \multicolumn{1}{c}{MIX1a}       & \multicolumn{1}{r}{MIX1b} 
& \multicolumn{1}{r}{MIX2a}       & \multicolumn{1}{r}{MIX2b} 
& \multicolumn{1}{r}{MIX3}        & \multicolumn{1}{r}{MIX4} \\
\hline 
\endfirsthead
\multicolumn{12}{c}%
{\tablename\ \thetable{} -- continued from previous page} \\
\hline\hline
\multicolumn{1}{l}{Model}         & \multicolumn{1}{r}{SM1}
& \multicolumn{1}{c}{SM3a($\%$)}  & \multicolumn{1}{c}{SM3b($\%$)}
& \multicolumn{1}{c}{SM4($\%$)}   & \multicolumn{1}{c}{PNM1($\%$)}
& \multicolumn{1}{c}{MIX1a}       & \multicolumn{1}{r}{MIX1b} 
& \multicolumn{1}{r}{MIX2a}       & \multicolumn{1}{r}{MIX2b} 
& \multicolumn{1}{r}{MIX3}        & \multicolumn{1}{r}{MIX4} \\
\hline 
\endhead
\hline
\multicolumn{12}{r}{{Continued on next page}} \\ 
\hline
\endfoot
\hline\hline
\endlastfoot
\multicolumn{12}{c}{linear finite range models} \\
\hline
H1 & $7.92$ & $100$U & $100$U & $100$U & $0$ & $-0.81$ & $-2.63$ & $0.41$ & $1.34$ & $-1.56$ & 
$-1.62$ \\ 
L1 & $9.89$ & $100$U & $100$U & $100$U & $0$ & $-1.66$ & $-4.33$ & $0.13$ & $0.83$ & $-0.61$ & 
$-1.69$ \\ 
L2 & $8.71$ & $100$U & $100$U & $100$U & $58.6$H & $-2.19$ & $-5.37$ & $-0.03$ & $0.55$ & $-0.28$ 
& $-1.69$ \\ 
L3 & $9.86$ & $100$U & $100$U & $100$U & $75.6$H & $-2.23$ & $-5.46$ & $-0.01$ & $0.58$ &  $-0.34$ 
& $-1.77$ \\ 
LBF & $8.85$ & $100$U & $100$U & $100$U & $0$ & $3.38$ & $5.77$ & $1.82$ & $3.88$ &   $-5.71$ & 
$-1.62$ \\ 
LHS & $7.95$ & $100$U & $100$U & $100$U & $0$ & $1.00$ & $1.00$ & $1.01$ & $2.42$ &   $-3.37$ & 
$-1.62$ \\ 
LW & $7.89$ & $100$U & $100$U & $100$U & $.4$L & $-1.58$ & $-4.16$ & $0.10$ & $0.78$ &   $-0.52$ & 
$-1.54$ \\
\end{longtable}
\pagebreak
\endgroup

\end{document}